\documentclass[10pt,twocolumn]{article}

\advance\topmargin by -1.00in
\usepackage{times} 
\usepackage{mathptm} 
\usepackage{graphicx} 
\usepackage{array} 
\usepackage{algorithm} 
\usepackage{url}

\begin{document}

\title
{Data Allocation in a Heterogeneous Disk Array - HDA \\
with Multiple RAID Levels for Database Applications}

\date{\vspace{-5ex}}
\vspace{-3mm}
\author{
\textit{Alexander Thomasian, IEEE Fellow\footnotemark[1]} \hspace{1mm}and Jun Xu
\footnotemark[2]
}
\maketitle

\footnotetext[1]
{Thomasian and Associates, 
17 Meadowbrook Rd, Pleasantville, NY 10570, US,
alexthomasian@gmail.com}

\footnotetext[2]
{Six Financial Information, New York, NY, xujun.sh@gmail.com.} 

	
\setcounter{page}{1}
\pagestyle{plain}
\thispagestyle{plain}


\begin{abstract}  
We consider the allocation of Virtual Arrays (VAs) in a Heterogeneous Disk Array (HDA).
Each VA holds groups of related objects and datasets such as files, relational tables,
which has similar performance and availability characteristics.
We evaluate single-pass data allocation methods for HDA using a synthetic stream of allocation requests,
where each VA is characterized by its RAID level, disk loads and space requirements. 
The goal is to maximize the number of allocated VAs and maintain high disk bandwidth and capacity utilization, 
while balancing disk loads. 
Although only RAID1 (basic mirroring) and RAID5 (rotated parity arrays) are considered in the experimental study, 
we develop the analysis required to estimate disk loads for other RAID levels. 
Since VA loads vary significantly over time, the VA allocation is carried out at the peak load period, 
while ensuring that disk bandwidth is not exceeded at other high load periods. 
Experimental results with a synthetic stream of allocation requests show 
that allocation methods minimizing the maximum disk bandwidth and capacity utilization 
or their variance across all disks yield the maximum number of allocated VAs. 
HDA saves disk bandwidth, 
since a single RAID level accommodating the most stringent availability requirements 
for a small subset of objects would incur an unnecessarily high overhead 
for updating check blocks or data replicas for all objects. 
The number of allocated VAs can be increased by adopting the clustered RAID5 paradigm, 
which exploits the tradeoff between redundancy and bandwidth utilization. 
Since rebuild can be carried out at the level of individual VAs, 
prioritizing rebuild of VAs with higher access rates can improve overall performance.
\end{abstract}

{\bf Keywords:} 
Storage systems, 
RAID, 
load balancing, 
data allocation, 
bin packing,
mirrored disks, 
rotated parity arrays, 
clustered RAID, 
bin-packing, 
queueing analysis, 
reliability analysis.

\section{INTRODUCTION}\label{sec:intro}

The {\it Redundant Array of Independent Disks (RAID)} paradigm \cite{Che+94}
ensures high availability as a solution to the high cost of downtime due to disk failures,
as quantified in Figure 1.3 in \cite{HePa07}.      
Replication and erasure coding are the two main methods to deal with disk failures 
and {\it Latent Sector Errors (LSEs)}, 
which result in unreadable disk blocks \cite{ThBl09}.
RAID level 0 (RAID0) added to the original five level RAID classification 
implements striping but provides no redundancy \cite{Che+94}.
Striping partitions large files into strips allocated in round-robin manner 
across the disks of the array with the intention of balancing disk loads.

RAID arrays were classified as {\it $k$-Disk Failure Tolerant ($k$DFT)} in \cite{ThBl09}.
{\it Maximum Distance Separable (MDS)} erasure coding requires $k$ check disks,
which is the minimum, to attain $k$DFT with RAID level $(4+k), k \geq 1$, 
so that the redundancy level with $N$ disks is $k/N$ \cite{ThBl09}.
This is much lower than $100 k\%$ redundancy ]of $k+1$-way replication.

RAID5 which tolerates single disk failures was extended to RAID6 to prevent unsuccessful rebuilds, 
which eventually lead to data loss, mainly due to LSEs rather than rare disk failures \cite{ThBl09}.
RAID6 utilizes two check strips per stripe according to {\it Reed-Solomon (RS)} coding, 
or computationally less costly parity coding in the case of EVENODD and RDP \cite{ThBl09}.
The disk access cost, which is the main consideration in this study, is the same in all three cases.
There are also 3DFTs which are MDS \cite{ThBl09}.  
The {\it Mean Time to Data Loss (MTTDL)} \cite{Che+94},
is a single measure used to quantify RAID resilience to failure,  
as determined by disk failure rate and the frequency of LSEs,
but asymptotic reliability analysis \cite{Thom06c} is used in Section~\ref{sec:ARA}. 
The failures of other components, such as the {\it Disk Array Controllers (DACs)} 
can be masked with internal redundancy such as duplexing, but also coding methods \cite{ThTH12}. 

A RAID level accommodating the most stringent availability requirements for a small subset of datasets 
would incur an unnecessarily high overhead for updating check blocks or data replicas for all datasets,
since as noted below disk bandwidths constitute a bottleneck. 
We propose {\it Heterogeneous Disk Arrays (HDAs)}
with DACs that can emulate multiple RAID levels and disks which share data for these levels.


Computer installation rather than procuring multiple disk arrays with different RAID levels,
which are inadequately utilized, could consolidate their data in HDAs, with adequate access bandwidth and capacity.
Fewer boxes entail a lower purchasing cost, footprint, and power consumption.
We are mainly concerned with database applications 
such as {\it OnLine Transaction Processing (OLTP)}, which entails random disk accesses,
and data mining which entails sequential accesses.
This study is not concerned with multimedia applications, such as {\it Video-on-Demand (VoD)}, 
with specialized data storage allocation methods and real-time data delivery requirements,
where the performance metric is the number of data streams supported.

\begin{figure}[htb]
\centerline{
\begin{tabular}{c}
\includegraphics[scale=0.75]{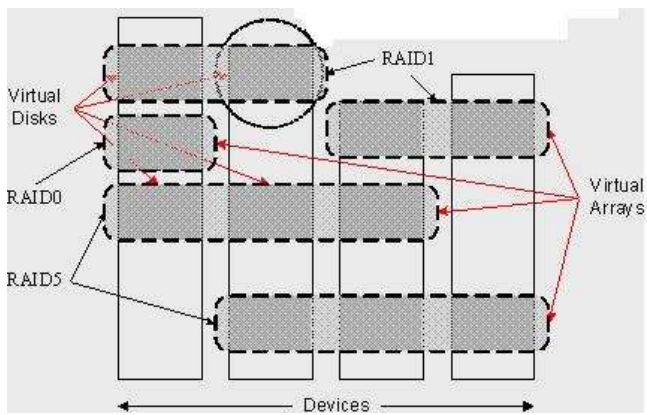}
\end{tabular}}
\caption{\label{fig:HDA} \small HDA with four disks
and five VAs with different RAID levels:
two RAID1 VAs with 2 disks each, 
two RAID5 VAs with 3 disks each,
and a single disk referred to as RAID0,
since it has no redundancy.} 
\vspace{1mm}
\end{figure}

{\it Virtual Arrays (VAs)} with RAID levels best suited for the respective application, 
are the units of data allocation in HDA.
A VA holds a collection of objects, datasets, such as files and relational tables,
which has similar performance and availability characteristics.
Figure \ref{fig:HDA} shows the sharing of disk space in an HDA with three RAID levels.
In fact only RAID levels RAID1 and RAID5 are considered in this study,
but we specify the formulas to estimate the load to allocate RAID0,                           
RAID$(4+k)$ for $k > 1 $, and clustered RAID arrays \cite{ThBl09}.
With a single failed disk in RAID5 the load on surviving disks is doubled,
which may result in overload in an HDA environment.
Clustered RAID sets the parity group size over which the parity is computed 
to a smaller value than the number of disks $(G<N)$,
which has the advantage of reduced load increase on surviving disks, i.e., $\alpha = (G-1)/(N-1)1<1$.


We assume that data is stored on {\it Hard Disk Drives (HDDs)},
which is the most popular storage medium for databases and file systems.
While disk capacities and transfer rates are increasing rapidly, 
the disk access time constitutes a bottleneck,
e.g., for ten milliseconds the access rate is limited to 100 I/Os per second (IOPs). 
The improvement in disk access time and hence disk bandwidth (accesses per second)
is quite limited for randomly placed disk blocks, which is typical for OLTP.
This is due to slow improvements in mechanical aspects of disk access mechanism \cite{JaNW08}.
The disk transfer rate for sequential data transfers 
is quite high due to high disk recording densities and increased RPMs (rotations per minute). 
DRAM and flash memories are more expensive per GB than disks,
so that they are utilized as disk caches, rather than HDD replacements.
Weaknesses associated with flash memories have led to a bleak outlook according to \cite{GrDS12}, 
but {\it Storage Class Memories (SCMs)} are expected to eventually replace HDDs due 
to their lower power consumption and footprint. 
{\it Non-Volatile Storage (NVS)} to hold dirty data is required to prevent data loss 
as in ordinary RAID arrays via {\it Uninterruptible Power Supply (UPS)} \cite{JaNW08}. 

The purchase cost of storage hardware,
which constitutes a significant fraction of the cost of computer installations,
is exceeded by {\it Total Cost of Ownership (TCO)}.
There is a need for tools to optimize storage bandwidth and capacity utilization, 
while providing satisfactory performance.
Several studies discussed in Section~\ref{sec:related}
have addressed the issue of optimal design of storage complexes. 
This study is more narrow in scope in that it is focused on improved storage utilization by data allocation.

The allocation of VAs in a disk array requires their load requirements.
This load can be obtained with various tools, 
such as IBM's {\it Resource Measurement Facility (RMF)},
\footnote{\url{http://www-03.ibm.com/systems/z/os/zos/features/rmf/}.}
and {\it Systems Management Facility (SMF)} allows a breakdown by workload, e.g., OLTP versus batch.
\footnote{\url{http://en.wikipedia.org/wiki/IBM_System_Management_Facilities}.}
%
%
It is difficult to specify VA loads without access to such data,
so that the alternate method of generating a synthetic stream of allocation requests for VAs 
is used in this study to compare the effectiveness of data allocation methods for HDA.
Each VA allocation request has a given RAID level, size, 
and its load in periods of high activity is known,
A VAs heavily loaded in one period may have a light load in another.
We carry out VA allocations based on the load in the heaviest period,
while keeping track of disk loads in other periods with heavy load
to ensure that disk utilizations are not exceeded in any period.


The primary goal of the allocation study is to maximize the number of allocated VAs,
with balancing disk bandwidth and capacity utilization serving as a secondary goal.
The former will improve disk response times.
VA parameters (load and size) are used to determine their width ($W$),
which is the number of {\it Virtual Disks (VDs)} required to materialize them.
The declustering ratio ($\alpha$) in clustered RAID5 \cite{Che+94,ThBl09}
is varied in Section~\ref{sec:craid} to take advantage 
of the tradeoff between disk capacity and bandwidth utilization in degraded mode. 
This results in an increased number of allocations when the system is disk bandwidth bound.

Optimization methods such as those discussed in \cite{PaRe02}
cannot be applied directly in this case for three reasons:                    
(1) this is not a straightforward 2-dimensional bin-packing problem,
because allocation requests are malleable,
e.g., clustered RAID5 provides a tradeoff between VA size 
and bandwidth or disk utilization requirements (in degraded mode).
(2) allocation requests become available 
and are assigned one at a time and not in batches.
The latter would allow optimization methods such as those discussed in \cite{PaRe02}.

The heuristic allocation method in this study formulates 
the allocation of VDs of a VA as vector-packing \cite{Hill94},   
which is equivalent to 2-dimensional bin-packing.
Small rectangles representing VD bandwidth and space requirement
are allocated into larger rectangles representing disk bandwidth and capacity.
Obviously, the smaller the bandwidth and space requirement per VD,
the larger the number of allocated VDs.

We consider single pass data allocation methods based on disk bandwidth and two new methods 
which take into account both bandwidth and capacity requirements.
The latter methods are more robust than the former.
A novel aspect of this study is that we take into account the load increase due to disk failures.
When the number of allocations is constrained by disk bandwidth, rather than disk capacity,   
we study the effect of adopting the clustered RAID5 paradigm,
since it offers a tradeoff between disk bandwidth and disk space requirements.


The paper is organized as follows.
Section \ref{sec:allocation} is concerned with data allocation in HDA.
%
Section \ref{sec:hda} on VA allocation in HDA first discusses data allocation in general
and then proceeds to discuss data allocation in normal and degraded modes.
%
Section \ref{sec:results} reports experimental results using a synthetic stream of allocation requests. 
%
A sensitivity study of the parameters used in the allocation study is reported in Section \ref{sec:sensitivity}.
Abbreviations are listed preceding the references.
Related work is discussed in the Appendix \ref{sec:related}.


\section
{\large DATA ALLOCATION BACKGROUND}\label{sec:allocation}

Section~\ref{sec:binpacking} provides the background for data allocation in HDA.
Section~\ref{sec:methods} specifies the methods used for allocating VDs.
In Section~\ref{sec:justification} we use a simplified analyses of 
HDA performance and reliability to justify its viability.

\vspace{-2mm}
\subsection
{Balancing Disk Allocations}\label{sec:binpacking} 

We model disks and allocation requests as vectors in two dimensions, 
where the x-axis is the disk access bandwidth or disk utilization
and the y-axis is its capacity \cite{Hill94}.
While very high capacity {\it Serial ATA (SATA)} disks 
with lower performance are available for archival storage,
we consider higher performance, smaller capacity SCSI disks \cite{AnDR03},
which are more appropriate for OLTP and database applications.

\begin{figure}[htb]
\centerline{
\begin{tabular}{c}
\includegraphics[scale=0.65,angle=270]{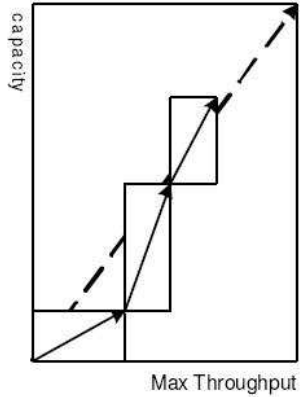}
\end{tabular}}
\caption{\small \label{fig:vectors} 
Allocation vectors for three VDs on a physical disk drive.
The x- and y-axis represent the disk bandwidth in processing accesses
to small randomly placed blocks and disk capacity utilizations.}
\end{figure}

We simplify the discussion by initially ignoring sequential accesses. 
A disk is then represented by the {\it Disk Vector} $\vec{D} = (X,C)$,
where $X$ denotes the maximum disk bandwidth for accessing small data blocks
or equivalently disk utilization in the case of sequential accesses,  
and $C$ the disk capacity, as shown in Figure~\ref{fig:vectors}.

Data allocation is modeled as vector addition.
Taking into account disk space is straightforward.
Disk bandwidth requirement is determined by assuming FCFS scheduling, average seek time,
which is based on one third of the maximum seek distance for uniform disk accesses,
and a mean latency equal to half a disk rotation time, plus the transfer time.      
This is a pessimistic assumption, since seek distances can be reduced using one of the following methods:
(i) Placing frequently accessed VAs centrally, i.e., the organ pipe organization.
(ii) Placing VAs accessed together on neighboring tracks,     
such as {\it Automatic Locality Improving Storage (ALIS)} \cite{JaNW08}.
(iii) {\it Shortest Access Time First (SATF)} disk scheduling significantly reduces 
disk positioning time with respect to FCFS under heavier disk loads \cite{JaNW08}.

The sum of all allocation vectors should not exceed the disk vector $\vec{D} = (X, C)$ in Figure \ref{fig:vectors}.
Full disk bandwidth and capacity utilization is attained when the end-point of allocation vectors reaches 
the end point of $\vec{D}$.
For heterogeneous disk drives $\vec{D}_n = (X_n, C_n)$, $1 \leq n \leq N$ 
have different values for $X_n$ and $C_n$ \cite{ThHa05}.
This will lead to unbalanced disk utilizations and higher disk access times on slower disks,
if the same stripe size is to be maintained across disks.

Disk allocation requests are at the level of VDs.
The resource requirements of VA$_i$'s VDs are specified as $\vec{d}_i = (x_i, c_i)$, 
where $x_i$ is the access rate (to small blocks) and $c_i$ is the disk space requirement per VD. 
With the initialization $X_n^r = X_n$, $C_n^r = C_n$, $1 \leq n \leq N$,
as additional VDs are allocated the residual disk bandwidth and capacity is updated as: 
$X_n^r = X_n^r - x_i$ and $C_n^r = C_n ^r - c_i$.
A VD allocation at the $n^{th}$ disk is successful if $ X_n^r \geq 0$ and $C_n^r \geq 0$, 
i.e., both constraints are satisfied.

Disk utilization is a more robust metric than disk bandwidth,
since the disk may be involved in processing disk accesses with varying sizes,
including sequential accesses.
Let $U_n^x$ and $U_n^c$ denote the current utilization of the bandwidth and capacity of the $n^{th}$ disk in the array
They are initialized as $U_n^X = 0$, $U_n^C = 0 $, $ 1 \leq n \leq N$
and the allocations are taken into account as 
$U_n^x = U_n^x + u^x_i  $ and $ U_n^c = U_n^c + u^c_i $, 
where $u^x_i$ and $u^c_i$ are the utilization of disk access bandwidth and capacity by VA$_i$.
Given that ${\cal J}_n$ denotes the set of VD allocations at the $n^{th}$ disk, we should have: 

\vspace{-4mm}
\begin{eqnarray}\label{eq:loading}
U_n^x = \sum_{j \in {\cal J}_n} u^x_j < 1 \hspace{5mm} 
U_n^c = \sum_{j \in {\cal J}_n} u^c_j < 1 .
\end{eqnarray}
The workload is considered disk bandwidth/utilization 
or capacity bound depending on which limit is reached first.



\vspace{-2mm}
\subsection
{Allocation Methods Considered in this Study}\label{sec:methods}

Five bin-packing methods based on disk bandwidth utilization  
and two additional methods which also take into account disk capacities
are considered for allocating VDs on $N$ disks in this study.
It is needless to say that VDs belonging to the same VA 
need to be allocated on different disks.
It is assumed that VA allocations are carried out in a disk array with no failed disks.
Allocation on consecutive disks simplifies addressing.


\begin{description}

\vspace{-2mm}
\item
[Round-Robin] 
Allocate VDs beyond the last stopping point,
on consecutive disk drives with wrap-around modulo $N$.
This method equalizes the allocated disk space
provided RAID1/0 and RAID5 arrays adopt the same strip size,
but does not balance bandwidth or capacity utilizations.

\vspace{-2mm}
\item
[Random:] 
Allocate VDs on disk drives randomly,
until an unsuccessful allocation is encountered.
Further attempts rarely result in a significant number of additional VA allocations, 
so that this option is not pursued further in this study.
This method does not ensure balanced capacity and bandwidth utilizations.

\vspace{-2mm}
\item
[Best-Fit:] 
Select the disk with the minimum remaining bandwidth 
or the maximum disk utilization, which may satisfy this request.

\vspace{-2mm}
\item
[First-Fit:] 
Disks are considered in increasing order of their indices, starting with the first disk.
A VD is allocated on the first disk that can hold it.

\vspace{-2mm}
\item
[Worst-Fit:] 
Allocate requests on disks with minimum bandwidth utilization, 
provided that both constraints are satisfied.

Given the bandwidth and capacity utilizations for the $n^{th}$ disk 
$U^x_n $ and $U^c_n $ given by Equations~(\ref{eq:loading}), 
two more sophisticated allocation methods minimize the following objective functions:

\def\var{\mathop{\rm Var}}

\vspace{-2mm}
\item
[Min-F1:]
Allocate VDs on disk drives, such that the maximum disk bandwidth utilization
or capacity utilization at the disks is minimized.
Note that without the latter constraint this is equivalent to Worst-Fit.

\vspace{-4mm}
\begin{equation}\label{obj-1}
\nonumber 
F_1 = \max_{1 \le n \le N} \left\{ U^x_n, \beta U^c_n \right\} , 
\hspace{5mm} 
\end{equation}

\vspace{-2mm}
\item
[Min-F2:]
Minimize the weighted sum of the variances of disk access bandwidths and capacity utilizations.

\vspace{-4mm}
\begin{equation}\label{obj-2}
F_2 = \var_{1 \le n \le N}(U^x_n) + \beta \var_{1 \le n \leq N}(U^c_n), 
\end{equation}
$\var(.)$ is the variance computed over all $N$ disks and 
$ \beta \geq 0 $ is an emphasis factor for disk capacity utilization.

\end{description}



\vspace{-2mm}
\subsection{Analytic Justification for HDA}\label{sec:justification}

Different RAID arrays can be implemented separately
or by dedicating subsets of the disks to each RAID level, 
e.g., allocating $n<N$ disks to RAID1 and $N-n$ disks to RAID5 in a single disk array.
This approach provides less flexibility than HDA,
because the mixture and resource demands for different RAID levels are not known a priori,
so that this may result in a situation when one array is underutilized and the other overutilized.
There is potential for improving disk access time by sharing disk space,
as demonstrated in Section~\ref{sec:improved}.   
%
There is a small reduction in reliability, however, as shown in Section~{sec:reliability}.
Calculation of mean response in an HDA with a disk failure is illustrated in Section~\ref{sec:degradedResp}.


In evaluating HDA performance we postulate an M/M/1 queueing system \cite{Klei75}
with Poisson arrivals and exponential service times.
Given an arrival rate $\lambda$ and mean disk service time $\overline{x}_d$,
the mean disk response time is $R = \overline{x}_d / (1- \rho)$,
where $\rho = \lambda \overline{x}_d$ is the disk utilization factor. 

\vspace{-3mm}
\subsubsection{Improved Response Time Due to HDA}\label{sec:improved}

The arrival rate of read requests to RAID1 and RAID5 arrays is 
$\Lambda_{R1}$ and $\Lambda_{R5}$, respectively.
Two disk array configurations are considered with $N=8$ disks:

\begin{description} 

\vspace{-2mm}
\item
[${\cal C}_1$: Vertical Disk Partitioning:]
$n=2$ disks are dedicated to RAID1 and $N-n=6$ disks to RAID5.

\vspace{-2mm}
\item
[${\cal C}_2$: Horizontal Disk Partitioning:]
RAID1 is allocated over $N=8$ disks, occupying $2/N$ or 25\% of the space on each disk.
The remaining empty space on the $N$ disks is more than adequate 
to hold the RAID5 array allocated on six disks in ${\cal C}_1$,
since the wider RAID5 array allocates less space to parity.
\footnote{The RAID5 data (not parity) blocks are distributed over seven disks,  
utilizing 5/7 or 72\% of disk capacity, which is less than 75\% of available space.}

\end{description}

The more important mean response time for RAID1 with ${\cal C}_1$ is:
$R_{R1} = \overline{x}_d / (1- \rho)$, 
where the disk utilization is $\rho = (\Lambda_1 /2 ) \overline{x}_d$.
The mean response time for RAID5 with ${\cal C}_1$ is
$R_{R5} = \overline{x}_d/(1-\rho)$,
where $\rho = (\Lambda_{R5} / 6) \overline{x}_{disk}$,

For ${\cal C}_2$ there are two components to disk utilization,
since RAID1 and RAID5 disk arrays share disk space:
$\rho' = [(\Lambda_{R1} + \Lambda_{R5}) / N] \overline{x}_d$,
so that ${R'}_{R1}  = {R'}_{R5} = \overline{x}_d / (1- \rho')$.
If $ \rho' < \rho $ or $\Lambda_{R5} < (N/2-1)\Lambda_{R1}$ then $ {R'}_{R1} < R_{R1}$.
For $N=8$ ${\cal C}_2$ will improve the RAID1 response time 
with respect to ${\cal C}_1$ for $\Lambda_{R1} > \Lambda_{R5} / 3$.

RAID1 response times in ${\cal C}_2$ can be improved by processing 
its accesses at a higher priority than RAID5 accesses.
${R''}_{R1} = \overline{x}_d / (1- \rho'')$,
where $\rho''= (\Lambda_{R1} / N) \overline{x}_d$, 
since only the disk utilization due to RAID1 accesses affects ${R''}_{R1}$ \cite{Klei75}.
For ${\cal C}_1$ with $\rho= (\Lambda_{R1}/ 2) \overline{x}_{disk} = 0.8$, 
$R_{R1} = \overline{x}_d / (1-0.8) = 5 \overline{x}_d$ 
and for ${\cal C}_2$ ${\rho''} = n \times 0.8 / N = 0.2$ and
${R''}_{R1} = \overline{x}_d / (1-0.2) = 1.25 \overline{x}_d$, i.e., a 4-fold improvement.

\vspace{-2mm}
\subsubsection{The Effect of Disk Failures on Response Time}\label{sec:degradedResp}

We use the HDA configuration in Figure~\ref{fig:HDA2} 
to illustrate the increases in mean disk response times 
as a result of a single disk failure.
All disk requests are reads with mean disk service time $\overline{x}_d$ 
and an arrival rate to each VA is $2 \lambda$,
so that the disk utilization per VD is $\rho_{VD} = \lambda \overline{x}_d$.
The utilizations of the eight disks in normal mode are then
$\rho =3 \rho_{VD}$ and the mean response times $R = \overline{x}_d / (1- \rho)$.

\begin{figure}[htb]
\renewcommand\arraystretch{1}
\centerline{\small
\begin{tabular}{|c c c c c c c c|}
 \hline
 1 & 2 & 3 & 4 & 5 & 6 & 7 & 8 \\
\hline
 $A_1$ & $A_2$ & $B_1$ & $B_2$ & $C_1$ & $C_2$ & $D_1$ & $D_2$ \\
 \hline
 $H_2$ & $E_1$ & $E_2$ & $F_1$ & $F_2$ & $G_1$ & $G_2$ & $H_1$ \\
\hline
 $L_1$ & $L_2$ & $I_1$ & $I_2$ & $J_1$ & $J_2$ & $K_1$ & $K_2$  \\
\hline
\end{tabular}}
\caption{\label{fig:HDA2}\small 
An HDA with $N=8$ disks with twelve VAs with mirrored disk configurations.}
\end{figure}

When disk 3 fails then the mean response time 
at VAs unaffected by this failure remains the same as before.
We assume that read requests are routed uniformly to the two VDs of a VA 
even when the load of the two disks on which they reside are not balanced.
Due to the unavailability of $E_2$, 
the load on $E_1$ will be doubled and hence $\rho_2 = 4 \rho_{VD} $.
This increased load affects the mean response time of VAs arrays ($A_1,A_2$) and ($L_1,L_2$).
Due to the unavailability of $B_1$ and $I_1$ the load on $B_2$ and $I_2$ will be doubled,
so that $\rho_4 = 5 \rho_{VD}$.
This increase affects the mean response time for VA ($F_1,F_2$).

The mean response times at the disk level are 
$R_1 = R_5 = R_6 = R_7 = R_8 = R$,
$R_2 = \overline{x}_d / (1- \rho_2)$,
$R_4 = \overline{x}_d / (1 -\rho_4)$,
so that the mean response time in degraded mode is 
$R' = [ R_1+ (4/3) R_2 + (5/3) R_4 + R_5 + R_6 + R_7 + R_8 ] / 7$.
Given $\rho_{VD}=0.1$ the mean response time in normal mode is
$R= (10/7) \overline{x}_d \approx = 1.43 $.
In degraded mode
$R' = [ 1+ (4/3)(10/6) + (5/3)(10/5) +4 ] /7 \approx 1.5 \overline{x}_d$,
which is a small increase in mean disk response time.
The mean response times for VAs in degraded mode are:
$R_{A_1,A_2} = R_{L_1,L_2} = (R+R_2)/2$,
$R_{E_1} = R_2$,
$R_{B_2} = R_{I2} = R_4$,
$R_{F_1,F_2} = (R_4+R)/2$.

\vspace{-3mm}
\subsubsection{Asymptotic Reliability Analysis of HDA}\label{sec:ARA}

We next utilize the asymptotic reliability analysis in \cite{Thom06c}
to compare the reliability of ${\cal C}_1$ and ${\cal C}_2$ configurations.
The reliability of RAID1/0 with $p$ disk pairs is 
$R_{R1}(p) = [1- (1-r)^2]^p$,
while the reliability of a RAID5 with $w$ disks is 
$R_{R5}(w) = r^w + w (1-r) r^{w-1}$.
In what follows we assume that the disks are highly reliable,
so that $r=1-\epsilon$ with $\epsilon \ll 1$.

\vspace{-2mm}
\[
R_{{\cal C}_1} =
[ 1- (1-r)^2] \times [r^6+6r^5(1-r)] \approx 1 -16\epsilon^2.
\]

\vspace{-5mm}
\[
R_{{\cal C}_2} = 
\{ [1-(1-r)^2]^4 \} \times [r^8 +8r^7 (1-r)] 
\approx 1-32\epsilon^2.
\]
It follows that the ${\cal C}_1$ configuration is more reliable than ${\cal C}_2$,
since it tolerates two disk failures as long as long as they do not affect the same array: RAID1 or RAID5.
This reliability analysis is extensible to more complicated cases.

\section {HDA DATA ALLOCATION}\label{sec:hda}

This section is organized as follow.
In Section~ \ref{sec:requests} we specify the characteristics of the VA allocation requests.
In Sections \ref{sec:normal} and \ref{sec:degraded}
we provide analytic expressions for system load per VA in normal and degraded modes.
The former load estimates are used to obtain the widths of the VAs
and the latter to estimate disk bandwidth requirements.

\vspace{-2mm}
\subsection{Virtual Allocation Requests}\label{sec:requests}

We are concerned with allocating VAs in an HDA with $N$ identical disks. 
Requests to allocate VAs become available one at a time and 
are assigned consecutive index numbers.
VA allocations are processed in strict {\it First-Come, First-Served (FCFS)} order,
i.e., VA$_{i+1}$ is allocated after VA$_{i}$.
Each VA is associated with a certain RAID level,
which cannot be reassigned for higher efficiency. 
A VA allocation fails when not all of its VDs can be allocated due to bandwidth or capacity constraints.
The goal is to maximize the number of allocations with balanced disk loads.


The VA load is determined by the arrival rate of disk requests and their mean service time.
Disk I/O trace analysis generated by OLTP workloads  
has shown that disk accesses are to small randomly placed disk blocks.  
Database applications, such as query processing and data mining,
require sequential accesses to large data files.
In the following discussion VAs are specified as follows:

\begin{itemize}

\vspace{-2mm}
\item 
{\it The RAID $\ell$evel of a VA}. 
This is specified a priori as $\ell=1$ for RAID1 and $\ell=5$ for RAID5.

\vspace{-2mm}
\item
{\it Ratio of RAID1 versus RAID5 arrays}.
A fraction $f_1$ of all allocation requests are for RAID1 
and a fraction $f_5 = 1 - f_1$ are for RAID5.
These fractions may follow the observed pattern 
or are based on classification as in \cite{And+02,ThXu11}.
We also use the notation RAID5:RAID1=3:1 implying $f_5=0.75$.

\vspace{-2mm}
\item 
{\it VA size}. 
The size $V_i$ of VA$_i$ is generated according to a RAID level dependent distribution,
but as if the data was to be allocated as a RAID0 array (with no redundancy).
The effective sizes of VA$_i$, denoted by ${V'}_i$, as determined below, 
are utilized in the allocation process.
The metadata required for HDA is specified in \cite{ThHa05}, but is not discussed in this study, 
since space requirements for metadata are considered to be negligible with respect to data size. 
Given the critical nature of metadata, 
multiple copies of it need to be stored on disk for the sake of higher reliability,
but also cached for the sake of efficiency.

For RAID1 VAs the effective size is ${V'}_i =2 V_i$,
since the disk space requirement is doubled. 
For RAID0/5/6/7 VAs which are $k$DFT arrays with $k = 0/1/2/3$,
the effective size of the VAs is ${V'}_i = V_i (W_i / (W_i - k ))$. 
The VA width $W_i$ is determined based on the size of the VA and its load in normal mode 
(see Equation (\ref{eq:width}) in Section \ref{sec:normal}).
${V'}_i$ in clustered RAID is discussed in Section \ref{sec:craid}. 
The same strip size may be adopted in all allocations,
which reduces external disk apace fragmentation,
while increasing internal disk fragmentation.

\vspace{-2mm}
\item
{\it VA loads}. 
Disk accesses to small randomly placed disk blocks incur a disk service time dominated by positioning time.
Restricted to such accesses the disk load per VA can be specified by its bandwidth to access small blocks.
Disk utilization is a more robust metric which is applicable 
to accesses to small blocks and large sequential accesses. 
A certain fraction of disk utilization may be reserved for sequential accesses.

The arrival rate of requests to VA$_i$ is $\Lambda_i = V_i \kappa_{\ell}$, 
where $\kappa_{\ell}$ is the I/O intensity per GB for RAID level $\ell$.
The per GB rate for RAID1 is set to be higher than RAID5 ($\kappa_1 > \kappa_5$),
since RAID1 arrays tend to be used for high performance OLTP applications. 
These parameters are also varied in the experimental study 
to emulate I/O-bound, balanced, and capacity-bound workloads.

VA loads may vary significantly in different periods.
The load in the heaviest period is specified explicitly,
but the loads in different periods are taken into account in data allocation,
so that they are not exceeded.

\vspace{-2mm}
\item
{\it Read:Write Ratio:}
The fraction of read and write requests to VA$_i$ is denoted by $f_r$ and $f_w = 1- f_r$, respectively.
Read (resp. write) requests to the disks are misses from a read cache 
(resp. destages from an NVS cache).
We ignore the possibility of repeated updates of dirty blocks and the potential locality for destages,
so in effect we postulate a worst case scenario.


\end{itemize}

Disk access time is determined by disk drive characteristics.
An analytic method to compute the mean disk service time for random disk accesses
with FCFS scheduling in zoned HDDs is reported in \cite{ThFH07}.
The {\it Shortest Access Time First (SATF)} disk scheduling policy 
currently implemented in most disk drives significantly improve disk access time \cite{JaNW08} 
and results in a lower disk utilization. 
FCFS scheduling is considered in this study, since it provides the worst case disk utilization.


The average disk transfer rate is considered in this study,
since the disk transfer rate varies across disk tracks 
due to {\it Zoned Bit Recording (ZBR)} \cite{JaNW08},
which maintains approximately the same linear recording density across tracks,
so that the mean disk transfer rate is a weighted sum according to track capacities
assuming that the track access rate is proportional to track size.


HDA introduces an additional layer in accessing data,
since there is a mapping for virtual to physical disk addresses, which are specified by LUNs.
This mapping is carried out using the metadata specified in \cite{ThHa05},
which is in addition to directories specifying 
the files (tables and indexes) in the case of a relational DBMS.
The data is cached so that there is no additional delay.

\vspace{-2mm}
\subsection
{Estimating VA Widths Based on Load in Normal Mode}\label{sec:normal}

Logical read and write requests are processed as 
{\it Single Read (SR)} and {\it Single Write (SW)} accesses.
The mean service times for SR, SW, and also {\it Read-Modify-Write (RMW)} disk accesses 
are denoted by $\overline{x}_{SR}$, $\overline{x}_{SW}$, and $\overline{x}_{RMW}$, respectively,
These are derived in Section \ref{sec:assumptions}. 

The load across the $W_i$ VDs of RAID0 is the arrival rate of requests 
to the array multiplied by the mean disk service time, 
which is a weighted sum of the two request types (${\rho'}_i)$.
Divided by the number of VDs ($W_i$) yields the HDD utilization per VD:

\vspace{-3mm}
\begin{eqnarray}
{\rho'}_i = 
\Lambda_i [ f_r \overline{x}_{SR} + f_w \overline{x}_{SW} ], \hspace{5mm} \rho_i = {\rho'}_i / W_i.
\end{eqnarray}

The width of RAID1/0 VAs is set to $W_i=2$ in this study,
since VA sizes for such arrays are selected to be relatively small.
The per VD utilization is the sum of the loads for read and write requests.
The read load per VD is also given based on the assumption 
that read requests are distributed evenly over the two disks.
\vspace{-4mm}
\begin{eqnarray}
{\rho'}_i = \Lambda_i [ f_r  \overline{x}_{SR} + 2 f_w \overline{x}_{SW} ],
\hspace{5mm}
\rho_i= \frac{{\rho'}_i}{2} =
\Lambda_i [\frac{ f_r}{2} \overline{x}_{SR} + f_w \overline{x}_{SW} ].
\end{eqnarray}
The cost equations for $(k+1)$-way replication are simple extensions of RAID1 costs.

In the case of RAID5/6/7 the updating of data and check blocks is accomplished via RMW requests, 
which read, modify, and then write data and check blocks.
Alternatives for updating small data blocks with the RMW method are:

\begin{description}

\vspace{-3mm}
\item[Method A]  
Given a newly modified data block ($d_{new}$), the DAC first reads the old data block ($d_{old}$), 
if it is not already cached, and compute $d_{diff}= d_{old} \oplus d_{new}$. 
The check blocks for the $i^{th}$ block of a modified EVENODD code can be computed as follows:
$p_{diff} = d_{diff}$,
$q_{diff} = \alpha^i d_{diff}$, 
$r_{diff}= \alpha^{2i} d_{diff}$ for the $i^{th}$ strip \cite{PlHu13}.
The old check blocks $p_{old}$, $q_{old}$, $r_{old}$ are also read from disk
to compute
$p_{new} = p_{old} \oplus p_{diff}$,
$q_{new} = q_{old} \oplus q_{diff}$,
$r_{new} = r_{old} \oplus r_{diff}$
which in addition to $d_{new}$ are written to appropriate disks as SW requests.
A logical write request requires $k+1=2/3/4$ SRs to read the old blocks and as many SWs to write them.
The resulting VA load is:

\vspace{-7mm}
\begin{eqnarray}
{\rho'}_i = 
\Lambda_i [ f_r \overline{x}_{SR} + (k+1)  f_w (\overline{x}_{SR} + \overline{x}_{SW})]. 
\end{eqnarray}

\vspace{-3mm}
\item[Method B]
We postulate an XOR capability at the disks to compute $d_{diff}$ and modified check blocks.
The DAC sends $d_{new}$ to the data disk to compute $d_{diff}$,  
which is sent via the DAC to the check disks,
specifying the identity of the check block (P, Q, or R).
The check disks read $p_{old}$, $q_{old}$, and $r_{new}$,
compute $p_{new}$, $q_{new}$, and $r_{new}$,
by applying appropriate coefficients and write them after one disk rotation,
so that $\overline{x}_{RMW}= \overline{x}_{SR} + T_{rot}$,
where $T_{rot}$ is the disk rotation time.
The VA load using this method is:

\vspace{-7mm}
\begin{eqnarray}
{\rho'}_i = \Lambda_i [ f_r \overline{x}_{SR} + (k+1)  f_w \overline{x}_{RMW}].
\end{eqnarray}

\vspace{-3mm}
\item[Method C]
The DAC after receiving $d_{diff}$ from the data disk computes 
$p_{diff}$, $q_{diff}$, and $r_{diff}$,
which are then sent to appropriate check disks for RAID5/6/7 disk arrays.
The difference with Method B is that only an XOR capability is required at the disks,
but the disk access time remains the same.

\vspace{-3mm}
\item[\bf Method D:]
{\it Disk Architecture for Composite Operations (DACO)}
reduce the cost of processing RMW accesses via a specialized read/write head \cite{LiSh10}, 
where the write head is placed at a short distance 
following the read head on a single disk actuator \cite{JaNW08}
allowing sufficient delay for a just read data block 
to be XORed with the difference block, before it is overwritten.
DACO result in a great improvement in processing RMW requests,
since $\overline{x}_{RMW} \approx \overline{x}_{SW}$,
but is not considered further because of difficulties associated with its implementation. 

\end{description}


Method A is less susceptible to lost updates, termed write holes, than Method B and C,
since the DAC logs its operations to recover from failures, such as power loss.
Methods B and C incur the same disk load.
Method A substitutes the disk rotation time associated with 
Methods B and C with an SW request (seek and latency).
By issuing of SW requests after SR requests are completed
and if there are no intervening disk accesses then an extra seek is not required at check disks, 
but almost a full disk rotation is incurred, 
so that the first three methods incur approximately the same cost: 
$\overline{x}_{SR} + \overline{x}_{SW} \approx \overline{x}_{RMW}$.

The VA width to implement RAID0/5/6/7 arrays is determined 
by the maximum disk utilization allowed per VD on each disk ($\rho_{max}$)  
and a maximum capacity constraint per VD $(v_{max})$,
which is expressed as a fraction of all disk capacities.
$W_i$ is the maximum of these widths, less than $N$:

\vspace{-5mm}
\begin{eqnarray}\label{eq:width2}
W_i^{bandwidth} = \lceil {\rho'}_i / \rho_{max} \rceil ,  \hspace{5mm}
W_i^{capacity} = \lceil V_i / v_{max} \rceil + k.
\end{eqnarray}

\vspace{-7mm}
\begin{eqnarray}\label{eq:width}
W_i = min \left[ max \left( W_i^{bandwidth} , W_i^{capacity} \right), N  \right] .
\end{eqnarray}


Limiting the per VD bandwidth utilization to a small fraction 
of disk bandwidth spreads the load across multiple disks.
This also reduces the possibility of disk overload when VA loads are underestimated. 
Allocating a RAID5 VA across all $N$ disks maximizes parallelism for read accesses 
and minimizes the space dedicated to check blocks,
but has the disadvantage that if a single disk fails then the read load at $N-1$ disks is doubled. 
While for $W<N$ the per VD load increase is higher than the case $W=N$,
fewer disk drives are affected by a disk failure.
HDA is effectively a clustered RAID since the VA widths $W < N$.
VA widths are reported in conjunction 
with allocation studies of clustered RAID in Section~\ref{sec:craid}.


Sequential accesses can be handled similarly to accesses to small blocks with the difference
that the seek time for sequential requests is negligible compared to transfer time.
Latency is reduced due to {\it Zero Latency Read (ZLR)} capability,
i.e., the reading of a sector can start at any any sector inside the block \cite{JaNW08}.
The preemptive resume policy is applied to large sequential accesses
to ensure an acceptable response time for small accesses. 
Preemptions incur extra positioning time so preemptions should be allowed 
at a sufficiently high granularity for efficiency reasons.
For the sake of brevity we omit the effect of sequential accesses on disk utilization,
which can be handled by estimating the service time for accesses to variable size blocks.
The analysis can be extended to accesses spanning  strips over several disks.
%
%
Large sequential updates can be processed efficiently as full-stripe writes,    
if all the strips in a stripe are updated.
Otherwise if the majority of the strips in a stripe are updated, 
the {\it ReConstruct Write (RCW)} method is more appropriate \cite{ThBl09}. 

\vspace{-2mm}
\subsection{Load in Degraded Mode}\label{sec:degraded}

We carry out VA allocations in a manner that HDA disk bandwidths 
are not exceeded in degraded mode when a single disk fails.
One method is to first carry out the allocations in normal mode,
but then consider the effect of failures of the disks allocated to a VA one by one, 
to ensure that no disk affected by the failure is overloaded.
Carrying out allocations by this method is costly because of the required backtracking.
Assume $I$ VAs are allocated on $N$ disks and the failure of a disk results in overload.
Then $i$ allocated VAs should be removed till there is no overload with $I-i$ allocated VAs. 
A larger $I-i$ can be attainable by moving data around.

The following alternate method was used in this study.
We allocate VAs in degraded mode to start with, as if there is a single disk failure.
Only single disk failures are considered for $k$DFT VAs even when $k>1$,
since concurrent disk failures are rare,
a disk failure is dealt with by rebuilt before an additional disk fails,
and the additional check blocks are intended to deal with media failures due to LSEs, 
rather than disk failures \cite{ThBl09}.
LSEs can be dealt with an intradisk redundancy scheme,
which applies the RAID paradigm at the level of disk segments \cite{ThBl09}.

An HDA with a single disk failure is not expected to operate 
in degraded mode for lengthy periods of time,
since the rebuild can be carried at the level of individual VAs
and the rebuild process for highly active VAs can be prioritized to improve overall performance.
Rebuild failures can be associated with individual VAs.


The read load on the surviving disk in RAID1 is doubled when one of the two disks fails,
while the write load remains the same.
Both disks for VA$_i$ are allocated with increased loads.
We use RAID1/F1 to specify RAID1 arrays with a single failure.

\vspace{-5mm}
\begin{eqnarray}\label{loadDegraded}
\rho^{RAID1/F1}  = \Lambda_i ( f_r \overline{x}_{SR} +  f_w \overline{x}_{SW} ).
\end{eqnarray}
Alternative RAID1 data organizations where the data on each disk is distributed
over multiple disks resulting in a more balanced load is described in \cite{ThXu08}.

Disk utilizations for read and write requests in clustered RAID5 
with a single failure (RAID5/F1) and a parity group size $G \leq W$  
are given below following the discussion in \cite{Thom05b}.
The parity group size is the number of strips, 
which is protected by a single parity strip in RAID5.
Using $W = W_i$ to simplify the notation,
the arrival rate per disk in VA$_i$ is $\lambda_i = \Lambda_i / W$. 

\vspace{-4mm}
\begin{eqnarray}\label{loadDegradedCRAID5read}
\rho_{read}^{CRAID5/F1} = 
\lambda_{i} r_{i} (1+ \alpha)  \overline{x}_{SR},
\hspace{5mm} \alpha = \frac{G-1}{W-1} < 1.
\end{eqnarray}

\vspace{-7mm}
\begin{eqnarray}\label{loadDegradedCRAID5write}
\rho_{write}^{CRAID5/F1} = 
\frac{\lambda_{i}w_{i}}{W-1} 
\left[ 2(W-2)\overline{x}_{RMW} + 2\overline{x}_{SW} + (G-2)\overline{x}_{SR}\right].
\end{eqnarray}

As far as sequential accesses are concerned, 
in the case of reads a missing strip can be reconstructed 
if all surviving strips in the stripe or parity group are read.
In the case of writes, the writing of the parity strip on a broken disk can be bypassed.

\section{ALLOCATION EXPERIMENTS}\label{sec:results} 

This section is organized as follows.
Section \ref{sec:experiment} describes the experimental method used to evaluate 
the data allocation methods considered in this study.
Section \ref{sec:assumptions} describes the assumptions made in the experimental study.
Section \ref{sec:main} summarizes the results of the experimental study to compare the allocation methods.
Sensitivity of results with respect to parameter settings to determine the width of VAs
and the parameter $\beta$ for the Min-F1 and Min-F2 method are evaluated in Section \ref{sec:sensitivity}.
We then assess the effect of clustered RAID5 on the number of allocations in Section \ref{sec:craid}.

\vspace{-2mm}
\subsection{Description of the Experiment}\label{sec:experiment}

Experiments to study the efficiency of VA allocations in HDA is reported in this section.
The attributes of VA allocation requests are generated one at a time 
using pseudo-random numbers sampling from prespecified distributions.
Requests are processed in strict FCFS order by different VA allocation policies. 
We are interested in identifying the policy that maximizes the number of allocated VAs.
The runs are repeated multiple times and the average over these runs is reported.
The same sequence of pseudo-random numbers is used 
to generate the same sequence of requests for different allocation policies.
This variance reduction technique is used for increased simulation efficiency.
\footnote{\url{http://en.wikipedia.org/wiki/Variance_reduction}.}

In addition to the number HDA disks and their characteristics,
the input parameters for the experiment are:
(i) The fraction of RAID1 versus RAID5 allocation requests in the input stream.
(ii) The size and bandwidth requirements for VAs,
where the latter are sampled from different distributions for RAID1 and RAID5.
(iii) The fraction of read and write requests.
The requests are to small randomly placed disk blocks,
so that the disk service time is determined by positioning time.

The experiment to estimate the efficiency of VA allocation methods can be specified as follows.

\begin{enumerate}    

\vspace{-2mm}
\item
Initialization: $I_{R1}=I_{R5}=0$ and the VA index $i=1$. 

\vspace{-2mm} 
\item
Determine the RAID level $\ell$ for VA$_i$ according 
to the fraction of RAID1 ($\ell=1$) versus RAID5 ($\ell=5$) VAs,
generate a uniformly distributed random number $u \in (0,1)$.
If $u \leq f_1 $ then VA$_i$ is a RAID1 and otherwise it is a RAID5. 

\vspace{-2mm}
\item
Generate the size of VA$_i$, which is denoted by $V_i$, 
based on the size distribution for RAID level $\ell$.

\vspace{-2mm}
\item
Compute the access rate to VA$_i$: $ \Lambda_i = \kappa_{\ell} V_i $,
where $\kappa_\ell$ depends on $\ell$ and workload category: 
bandwidth-bound, capacity-bound, or balanced.
The load due to sequential accesses is expected to be proportional to $V_i$,
and remains the same in normal and degraded mode.                 

\vspace{-2mm}
\item
For RAID5 allocation requests determine the VA load according to Section~\ref{sec:normal}.
Use thresholds for maximum disk bandwidth ($\rho_{max})$ 
and capacity ($v_{max}$) using Equation~(\ref{eq:width}).
Set $W_i = 2$ for RAID1 VAs.

\vspace{-2mm}
\item
Calculate VA loads in degraded mode using the expressions in Section \ref{sec:degraded}.

\vspace{-2mm}
\item
Determine if all VDs of VA$_i$ can be allocated successfully 
satisfying disk space and bandwidth constraints, the latter in {\it degraded mode}. 
\footnote{All of the VDs of a  VA are allocated in degraded mode,
since it is not known a priori which VD will be placed on a disk that fails later.}
Ensure that the load in no other high load period is exceeded.
If so set $I_{R1\&R5} = I_{R1} + I_{R5}$ and stop.

\vspace{-2mm}
\item
Set $I_{R\ell} = I_{R\ell} +1$ based on the RAID level $\ell$ of the allocated VA$_i$.
Increment the disk bandwidth and capacity utilization or decrement 
the residual bandwidth and capacity for all allocation periods being considered,
set $i=i+1$, and return to Step 2.

\end{enumerate}

This experiment can be repeated for normal mode operation
using the disk utilizations in Section~\ref{sec:normal} for comparison purposes.
Allocation methods are ranked by the number of allocated VAs,
which is specified as $I_{R1\&R5} = I_{R1} + I_{R5}$, 
where $I_{R1}$ and $I_{R5}$ are counters for allocated RAID1 and RAID5 VAs.

VA loads may vary significantly over time, so that we have VAs sharing disk space,
which reach their peak load in different periods.
In this case the algorithm should be carried out using the load in the peak period, 
but ensuring that the disk load in no other high load period is exceeded.


\vspace{-2mm}
\subsection{Assumptions and Parameter Settings}\label{sec:assumptions}

We use the following parameter settings to run experiments.
An HDA with $N=12$ IBM 
model 18ES drives is considered.
The characteristics of these disk drives, 
which are extracted from the web site for the {\it Parallel Data Laboratory (PDL)} 
\footnote{\url{http://www.pdl.cmu.edu/DiskSim/diskspecs.shtml}.}
are summarized in Table~\ref{tab-hd-spec}. 
The relative efficiency of data allocation methods is expected to hold
due to the low rate of improvement in disk access times \cite{JaNW08}.

\begin{table}[htb] 
\centerline{ 
\begin{tabular}{|c|c|} \hline 
Model & IBM DNES-309170W \\ 
\hline%
Disk Capacity & $C=9.17$ GB \\ 
\hline %
Number of cylinders/zones & 11,474/11 \\ 
\hline %
Tracks per cylinder & 5 \\ 
\hline %
\hline %
Rotations per minute & 7200 RPM \\ 
\hline %
Disk rotation time & $T_{rot} = 8.33$ ms \\ 
\hline 
latency & $\overline{x}_{lat} \approx T_{rot} /2 $ \\ 
\hline 
Head settling time & $T_{h} = 0.14 $ ms \\ 
\hline %
Mean seek time & $\overline{x}_{seek} \approx 7.16$ ms \\ 
\hline 
Mean transfer time per blocks & 
$\overline{x}_{xfer} \approx 0.16$ ms \\ 
\hline Mean read time ($\overline{x}_{SR}$) & $ \overline{x}_{seek}+ 
\overline{x}_{lat} + \overline{x}_{xfer}$ \\ 
\hline Read time for RAID5 & $\overline{x}_{SR} \approx 11.49$ ms \\ 
\hline 
Disk write time & $\overline{x}_{SW}= \overline{x}_{SR}+T_h $ \\ 
\hline 
RMW time for RAID5 & $\overline{x}_{RMW}=\overline{x}_{SR}+T_{rot}$\\ 
\hline \end{tabular}} 
\caption{\label{tab-hd-spec}\small Characteristics of IBM 18ES disk drives.
The mean seek and transfer time are computed assuming random accesses to disk blocks
taking into account disk zoning.}
\end{table} 

The size of the nonredundant data stored in each VA is assumed 
to be exponentially distributed with a mean $\overline{V}_1= 256$ MB for RAID1 
and $\overline{V}_5 = 768$ MB for RAID5 VAs. 
The RAID sizes obtained by sampling are rounded up to multiples of 256 KB,
which can be placed in strips of this size.
As noted earlier space requirements for metadata are ignored in this study.

We set $v_{max}=1/50$  of the capacity of all disks 
and $\rho_{max}= 1/20$ the bandwidth of each disk. 
We set $\beta =1$ based on the sensitivity analysis in Section~\ref{sec:sensitivity} 
for the Min-F1 and Min-F2 allocation methods. 
The access rate for RAID1 VAs is set to be ten times higher than the rate for RAID5 VAs, 
i.e., $\kappa_1 = 10 \kappa_5$.
While RAID1 arrays are three times smaller on the average,
the rate of accesses to RAID1 arrays 
is set to be 3.3 times higher than RAID5 on the average.
We consider three VA workloads with different access rates for RAID5 VAs in normal mode.
Bandwidth-Bound: $\kappa_5 = 8.5$, Balanced: $\kappa_5 = 3.3$, 
Capacity-Bound: $\kappa_5 = 2.1$ accesses/second per GB.

We consider three cases for the composition of read and write requests:
(i) all requests are reads ($f_r=1$),
(ii) 75\% of requests are reads ($f_r=0.75$), 
(iii) 50\% of requests are reads ($f_r=0.5$).
Parameters used in the experimental study are summarized in Table~\ref{tab-exp}.

\begin{table}[htb]
\footnotesize
\centerline{ 
\begin{tabular}{|c|c|}                            \hline %
Number of disks            & $N=12$          \\   \hline 
RAID1/5 allocations fraction & $f_1$, $f_5=1-f_1$           \\   \hline
Per GB rate to RAID5/RAID1       & $\kappa_5$, $kappa1_1=10\kappa_5$      \\   \hline
Mean RAID1 size            & $\overline{V}_1= 256 MB$    \\ \hline
Mean RAID5 size            & $\overline{V}_5= 768 MB$    \\ \hline
Load for VA$_i$            & ${\rho'}_i$                  \\   \hline
Fraction of reads/writes   & $f_r$/$F_w=1-f_r$             \\   \hline
Maximum bandwidth per VD   & $\rho_{max}=1/20=0.005$    \\   \hline
Maximum capacity per VD    & $v_{max}=1/50=4\%$ (al disks)      \\   \hline
RAID5 VA width             & $W$             \\   \hline
Parity group size          & $G$             \\   \hline
Declustering ratio  & $\alpha = \frac{G-1}{W-1}$  \\   \hline
Capacity emphasis factor   & $\beta \geq 0 $      \\   \hline 
\end{tabular}}
\caption{\label{tab-exp}\small Parameters used in the experimental study.}
\end{table}

\vspace{-2mm}
\subsection
{Comparison of Allocation Methods}\label{sec:main}

\begin{table*}[t]
\centerline{\small
\begin{tabular}{|c||c|c|c|c||c|c|c|c||c|c|c|c|}
\hline
&  \multicolumn{4}{c||}{Bandwidth-Bound} & \multicolumn{4}{c||}{Balanced} & \multicolumn{4}{c|}{Capacity-Bound}\\
\cline{2-13}
& & \multicolumn{3}{c||}{Allocations} & &  \multicolumn{3}{c||}{Allocations}& & \multicolumn{3}{c|}{Allocations}\\
\cline{2-13}
    Method & {Best} & R1 & R5 & R1\&R5 & {Best} & R1 & R5 & R1\&R5 & {Best} & R1 & R5 & R1 \& R5\\
\hline\hline
Min-F1  &   80  &   47.1    &   142.4   &   189.6   & 90 &   57.2 &   173.2   &   
230.4   &   90  &   69.2    & 206.7 & 275.9   \\
\hline
Min-F2  &   78  &   46.4    &   142.3 &   188.6 &   76  & 56.2 & 172.4   &   
228.6   &   86  & 68.2    & 205.6   & 273.8   \\
\hline
Worst-Fit   &   66  & 45.2    & 137.6 &   182.8 &   15  & 48.1 &   150.4   & 
198.5   &   9 &   61.3 &   183.6 &   244.9   \\
\hline
Best-Fit    &   63  & 41.1    & 123.1   & 164.2   &   12  & 44.3 & 135.3   & 
179.6   &   8   & 57.4    & 178.3   &   235.7 \\
\hline
Round-Robin &   17  &   33.3    &   102.3   &   135.6   & 10 & 44.4 &   133.4   
&   177.8   &   8   &   63.2    & 190.1   & 253.3   \\
\hline
First-Fit   &   13  &   33.1    & 100.4   & 133.5 &   0   & 35.1 & 103.0   &   
138.1   & 0   &   38.1 &   115.3 &   153.5 \\
\hline
Random  &   13  & 29.1    & 90.3    &   119.4 &   6   & 33.0 & 107.3 & 140.3   &   2 &   57.2    & 172.2   &   229.3\\
\hline
\end{tabular}}           
\caption{\label{tab:resultsNormal-1-0} \small Comparison of
the allocation methods with RAID5:RAID1=3:1, and $r=1$ in normal mode.} 

\vspace{2mm}

\centerline{\small
\begin{tabular}{|c||c|c|c|c||c|c|c|c||c|c|c|c|}
\hline
&  \multicolumn{4}{c||}{Bandwidth-Bound} & \multicolumn{4}{c||}{Balanced} & 
\multicolumn{4}{c|}{Capacity-Bound}\\
\cline{2-13}
& & \multicolumn{3}{c||}{Allocations} & &  \multicolumn{3}{c||}{Allocations}& 
& \multicolumn{3}{c|}{Allocations}\\
\cline{2-13}
Method & {Best} & R1 & R5 & R1 \& R5 & {Best} & R1 & R5 & R1 \& R5 
& {Best} & R1 & R5 & R1 \& R5\\
\hline\hline 
Min-F1  &   78  &   41.4    &   125.1   &   166.5   & 87  &   50.2&   152.1   &   
202.3   &   89  &   68.5    &   204.7   &   273.2   \\
\hline
Min-F2  &   75  &   40.7    &   124.9 &   165.7   &   73  & 49.3&   151.4   &   
200.7   &   85  &   67.5    &   203.6   &   271.1   \\
\hline
Worst-Fit   &   63  & 39.7    &   120.8   & 160.5   &   14  & 42.3&   132.1   & 
174.4   &   9   &   60.7    &   181.7   &   242.4   \\
\hline
Best-Fit    &   61  &   36.1    & 108.1   &   144.2   & 12  & 38.9 & 118.8   &   
157.8   &   8   &   56.8    &   176.6   &   233.4   \\
\hline
Round-Robin &   16  &   29.2    &   89.8    &   119.1   & 10  &   39.0    &   
117.2   &   156.2   &   8   &   62.6    &   188.2   &   250.8   \\
\hline
First-Fit   &   13  &   29.1    & 88.2    & 117.2   &   0   &   30.8    &   
90.5    &   121.3   & 0   &   37.7    &   114.2   &   151.9   \\
\hline
Random  & 12  & 25.5 & 79.3 & 104.8   &  6   &  29.0  & 94.3  & 123.2   &   2   &   
56.6    &   170.5   &   227.1   \\
\hline
\end{tabular}}
\caption{\label{tab:resultsNormal-3-1} \small Comparison of
the allocation methods with RAID5:RAID1=3:1, and $r=0.75$ in normal mode.}

\vspace{2mm}

\centerline{\small
\begin{tabular}{|c||c|c|c|c||c|c|c|c||c|c|c|c|}
\hline
&  \multicolumn{4}{c||}{Bandwidth-Bound} & \multicolumn{4}{c||}{Balanced} & 
\multicolumn{4}{c|}{Capacity-Bound}\\
\cline{2-13}
& & \multicolumn{3}{c||}{Allocations} & &  \multicolumn{3}{c||}
{Allocations}& & \multicolumn{3}{c|}{Allocations}\\
\cline{2-13}
Method & {Best} & R1 & R5 & R1 \& R5 & {Best} 
& R1 & R5 & R1 \& R5 & {Best} & R1 & R5 & R1 \& R5\\
\hline\hline
Min-F1  &   75  &   32.4    &   97.9    &   130.3   & 84  &   39.3    &   119.1   
&   158.4   &   89  &   68.2    &   203.7   &   271.9   \\
\hline
Min-F2  &   73  &   31.9    &   97.8 &   129.7   &   70  & 38.6    & 118.5   &   
157.2   &   85  &   67.2    &   202.6   &   269.8   \\
\hline
Worst-Fit   &   61  & 31.1    &   94.6    & 125.7 &   14  & 33.1 &   103.4   & 
136.5   &   9   &   60.4    &   180.8   &   241.2   \\
\hline
Best-Fit    &   59  &   28.3    & 84.6    & 112.9   & 12  &   30.5 &   93.0    
&   123.5   &   8   &   56.6    &   175.7   &   232.3   \\
\hline
Round-Robin &   16  &   22.9    &   70.3    &   93.2    & 10  &   30.5    &   
91.7    &   122.3   &   8   &   62.3    &   187.3   &   249.5   \\
\hline
First-Fit   &   12  &   22.8    & 69.0    & 91.8    &   0   &   24.1    &   
70.8    &   95.0    & 0   &   37.6    &   113.6   &   151.2   \\
\hline
Random  &   12  & 20.0    & 62.1    &   82.1    &   6   &   22.7    &   73.8    & 
96.5    &   2   &   56.3    &   169.6   &   225.9   \\
\hline
\end{tabular}}
\caption{\label{tab:resultsNormal-1-1} \small Comparison of
the allocation methods with RAID5:RAID1=3:1, and $r=0.5$ in normal mode.}
\end{table*}

\begin{table*}[t]
\centerline{\small
\begin{tabular}{|c||c|c|c|c||c|c|c|c||c|c|c|c|}
\hline
&  \multicolumn{4}{c||}{Bandwidth-Bound} & \multicolumn{4}{c||}{Balanced} & 
\multicolumn{4}{c|}{Capacity-Bound}\\
\cline{2-13}
& & \multicolumn{3}{c||}{Allocations} & &  \multicolumn{3}{c||}{Allocations}& 
& \multicolumn{3}{c|}{Allocations}\\
\cline{2-13}
    Method & {Best} & R1 & R5 
& R1 \& R5 & {Best} & R1 & R5 & R1 \& R5 & {Best} & R1 & R5 & R1 \& R5\\
\hline\hline
Min-F1  &   71  &   23.2    &   72.0    &   95.2    &   87  &   28.2    &   
84.6    &   112.8   &   87  &   34.2    &   105.3   &   139.5   \\
\hline
Min-F2  &   71  &   23.2    &   72.0    &   95.2    &   73  &   27.3    &   
84.6    &   111.9   &   83  &   33.4    &   102.3   &   135.7   \\
\hline
Worst-Fit   &   56  &   21.8    &   71.6    &   93.4    &   14  &   23.7    &   
75.2    &   98.9    &   9   &   29.7    &   93.4    &   123.1   \\
\hline
Best-Fit    &   54  &   20.3    &   68.7    &   89.0    &   12  &   22.0    &   
70.1    &   92.1    &   8   &   27.6    &   90.2    &   117.8   \\
\hline
Round-Robin &   19  &   16.4    &   50.6    &   67.0    &   10  &   21.1    &   
66.4    &   87.5    &   8   &   30.8    &   95.1    &   125.9   \\
\hline
First-Fit   &   10  &   16.0    &   48.5    &   64.5    &   0   &   18.7    &   
56.8    &   75.5    &   0   &   19.3    &   59.7    &   79.0    \\
\hline
Random  &   10  &   13.7    &   47.2    &   60.9    &   6   &   20.2    &   61.7    &   
81.9    &   2   &   28.0    &   86.9    &   114.9   \\
\hline
\end{tabular}}

\caption{\label{tab:resultsDegraded} \small Comparison of the
allocation methods with RAID5:RAID1=3:1, and $r=1$ in degraded mode. 
}

\vspace{2mm}

\centerline{\small
\begin{tabular}{|c||c|c|c|c||c|c|c|c||c|c|c|c|}
\hline
&  \multicolumn{4}{c||}{Bandwidth-Bound} & \multicolumn{4}{c||}{Balanced} & 
\multicolumn{4}{c|}{Capacity-Bound}\\
\cline{2-13}
& & \multicolumn{3}{c||}{Allocations} & &  \multicolumn{3}{c||}{Allocations}& 
& \multicolumn{3}{c|}{Allocations}\\
\cline{2-13}
    Method & {Best} & R1 & R5 
& R1 \& R5 & {Best} & R1 & R5 & R1 \& R5 & {Best} & R1 & R5 & R1 \& R5\\
\hline\hline
Min-F1  &   82  &   17.3    &   64.0    &   81.3    &   98  &   23.2    &   
72.6    &   95.8    &   80  &   33.0    &   104.0   &   137.0   \\
\hline
Min-F2  &   75  &   17.3    &   64.0    &   81.3    &   69  &   23.1    &   
73.5    &   96.6    &   83  &   32.1    &   101.0   &   133.1   \\
\hline
Worst-Fit   &   66  &   16.5    &   59.6    &   76.1    &   10  &   21.3    &   
66.9    &   88.2    &   7   &   29.0    &   92.0    &   121.0   \\
\hline
Best-Fit    &   63  &   15.0    &   58.8    &   73.8    &   9   &   19.6    &   
62.4    &   82.0    &   6   &   27.1    &   89.0    &   116.1   \\
\hline
Round-Robin &   12  &   12.8    &   45.6    &   58.4    &   11  &   16.8    &   
56.6    &   73.4    &   6   &   30.0    &   92.1    &   122.1   \\
\hline
First-Fit   &   16  &   13.7    &   44.7    &   58.5    &   0   &   17.9    &   
53.8    &   71.7    &   0   &   18.1    &   58.0    &   76.1    \\
\hline
Random  &   12  &   14.0    &   38.7    &   52.7    &   7   &   17.8    &   
53.6    &   71.4    &   2   &   27.0    &   85.0    &   112.1   \\
\hline
\end{tabular}}
\caption{\small \label{tab:resultsDegraded-r-075} \small Comparison
of the allocation methods with RAID5:RAID1=3:1 and $r=0.75$ in degraded mode. 
}
\vspace{3mm}

\centerline{\small
\begin{tabular}{|c||c|c|c|c||c|c|c|c||c|c|c|c|}
\hline
&  \multicolumn{4}{c||}{Bandwidth-Bound} & \multicolumn{4}{c||}{Balanced} & 
\multicolumn{4}{c|}{Capacity-Bound}\\
\cline{2-13}
& & \multicolumn{3}{c||}{Allocations} & &  \multicolumn{3}{c||}{Allocations}& 
& \multicolumn{3}{c|}{Allocations}\\
\cline{2-13}
    Method & {Best} & R1 & R5 & R1 \& R5 
& {Best} & R1 & R5 & R1 \& R5 & {Best} & R1 & R5 & R1 \& R5\\
\hline\hline
Min-F1  &   78  &   15.3    &   57.6    &   72.9    &   98  &   21.7    &   
66.6    &   88.3    &   84  &   33.0    &   104.0   &   137.0   \\
\hline
Min-F2  &   75  &   15.3    &   55.7    &   71.1    &   71  &   20.9    &   
67.4    &   88.3    &   90  &   32.1    &   101.0   &   133.1   \\
\hline
Worst-Fit   &   63  &   14.7    &   51.9    &   66.6    &   6   &   19.9    &   
63.8    &   83.7    &   3   &   29.0    &   92.0    &   121.0   \\
\hline
Best-Fit    &   61  &   13.3    &   51.2    &   64.5    &   5   &   18.2    &   
59.5    &   77.7    &   3   &   27.1    &   89.0    &   116.1   \\
\hline
Round-Robin &   13  &   10.7    &   39.4    &   50.1    &   10  &   14.0    &   
47.5    &   68.7    &   6   &   30.0    &   92.1    &   122.1   \\
\hline
First-Fit   &   16  &   12.0    &   39.6    &   51.6    &   0   &   16.9    &   
51.8    &   61.5    &   0   &   18.1    &   58.0    &   76.1    \\
\hline
Random  &   12  &   11.2    &   32.9    &   44.1    &   3   &   15.2    &   47.2    &   
62.4    &   2   &   26.1    &   83.2    &   109.3   \\
\hline
\end{tabular}}
\caption{\label{tab:resultsDegraded-r-05} \small Comparison
of the allocation methods with RAID5:RAID1=3:1 and $r=0.5$ in degraded mode. 
}
\end{table*}

The number of allocated RAID1 and RAID5 VAs 
is the key performance metric in evaluating data allocation methods.
The number of allocated VAs in the two categories follows the fractions of requests in the input stream,
since the allocations are carried out in FCFS order.

Tables \ref{tab:resultsNormal-1-0},
\ref{tab:resultsNormal-3-1}, 
and \ref{tab:resultsNormal-1-1} 
are based on 100\%, 75\%, and 50\% read requests in normal mode, 
while Tables \ref{tab:resultsDegraded}, 
\ref{tab:resultsDegraded-r-075}, and
\ref{tab:resultsDegraded-r-05} 
are based on the same fractions of read requests in degraded mode.
The three cases: bandwidth-bound, balanced, and capacity-bound workloads are considered.
The same pseudo-random sequence was used in experiments to generate 
the same synthetic sequence of allocation requests for a fair comparison.

The allocation experiments were repeated one hundred times,
so as to obtain the average number of allocated VAs over these iterations.
Increasing the number of iterations yielded indistinguishable results,
so lengthier experiments are not reported here.
The key metric in the comparison is the number of times an allocation method performed best,
i.e., allocated the most VAs.
It follows from the tables that there were many ties,
with more than one method providing the best allocation.
The following conclusions can be drawn from the tables.

\begin{enumerate}

\item
The number of VAs allocated in normal mode is almost 
double the number of those in degraded mode for $r=1$.
This is because the load on surviving disks in degraded mode 
is approximately double the load in normal mode (for both RAID1 and RAID5 arrays).   

\item
Min-F1 and Min-F2 are consistently the best methods
in terms of the number of allocations in all configurations,
although Min-F1 outperforms Min-F2.
Both disk bandwidth and capacity need to be considered for robust resource allocation.

\item
Worst-Fit and Best-Fit methods provide good allocations 
for bandwidth-bound workloads in normal and degraded modes.
The number of allocations is lower than Min-F1 and Min-F2, however.

\item
First-Fit, Random, and Round-Robin 
are the worst among all allocation methods considered,
since they do not optimize bandwidth or capacity allocation.

\end{enumerate}

Table~\ref{OneRunMixedBw}, 
provides the comparison of average bandwidth and capacity utilization 
for all disks and number of RAID1 and RAID5 VAs allocated with RAID5:RAID1=3:1, 
$r=1$ in degraded mode for Min-F1, Min-F2, 
and the less efficient Round-Robin allocation methods, for bandwidth-bound workload.

The following observations can be made:
Min-F1 and Min-F2 attain high disk bandwidth utilization for bandwidth-bound and balanced workloads,
while disk capacity utilizations are low for bandwidth-bound workloads,
since the bandwidth bound is reached first.
For capacity-bound allocations Min-F1 and Min-F2 minimize the variation of disk utilizations
and this results in an increase in the number of VA allocations. 
Poor allocation methods have high standard deviations
for disk bandwidth and capacity utilization.

\begin{table}[htp]
\centerline{ \small
\begin{tabular}{|c||c|c||c|c||c|c|c|}
\hline
& \multicolumn{2}{c||}
{Bandwidth} &  \multicolumn{2}{c||}
{Capacity)}& \multicolumn{3}{c|}{No. of VAs}\\
\cline{2-8}
    Method & Avg & Std & Avg & Std & R1 & R5 & Total\\
\hline\hline
    Min-F1 & 90.7 & 1.6 & 51.3 & 2.6 & 23 & 72 & 95\\
\hline
    Min-F2 & 90.4 & 2.3 & 51.3 & 2.2 & 23 & 72 & 95\\
\hline
    RR & 63.8 & 20.1 & 36.8 & 8.4 & 16 & 51 & 67\\
\hline
\end{tabular}}
\caption{\small \label{OneRunMixedBw} \small 
Disk utilizations after allocations with RAID5:RAID1=3:1, $r=1$, 
and bandwidth-bound workload in degraded mode.}
\end{table}



\begin{figure}[htp]
\includegraphics[scale=0.7]{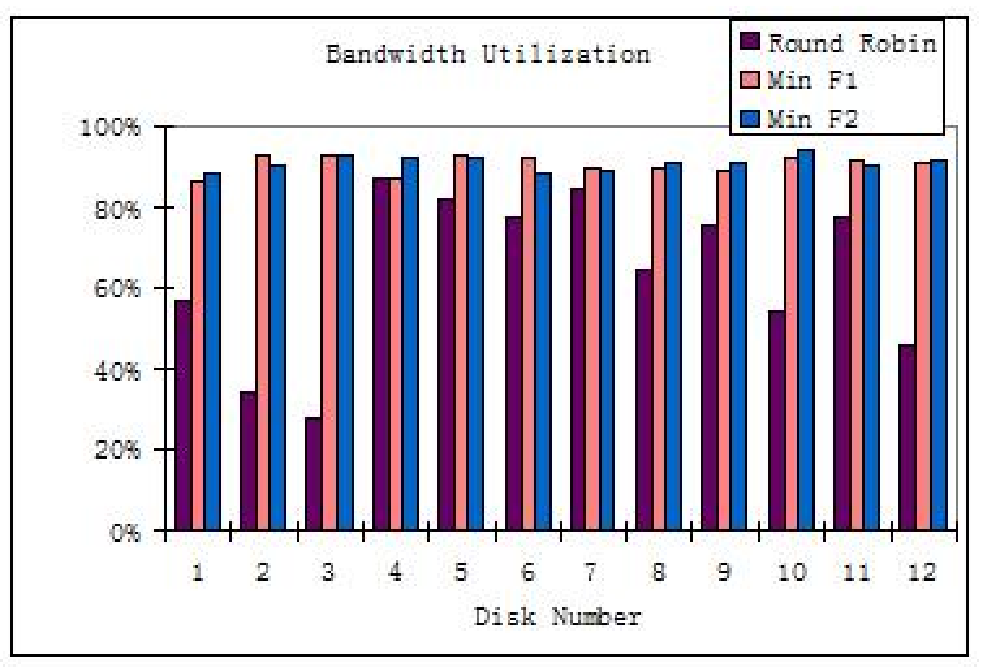}   
\includegraphics[scale=0.7]{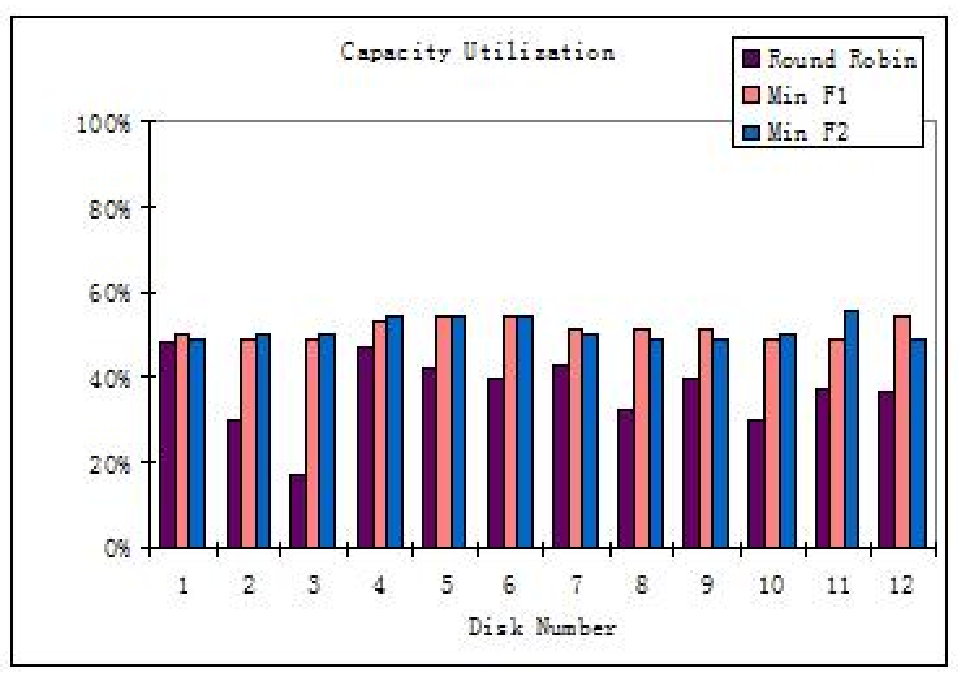}
\caption{\small \label{alloc-util-bw}\small 
Disk bandwidth and capacity utilizations with RAID5:RAID1=3:1, $r=1$, 
and bandwidth-bound workload in degraded mode.}
\end{figure}



Disk bandwidth and capacity utilizations with bandwidth-bound workloads 
for Min-F1, Min-F2, and Round-Robin allocation methods
are shown in Figures~\ref{alloc-util-bw}                 
It follows that for Min-F1 and Min-F2 all disks 
are saturated roughly equally as far as disk bandwidth or capacity
or both are concerned for bandwidth-bound, capacity-bound, and balanced workloads, respectively.
Little or no improvement in allocations is therefore expected
with more sophisticated allocation methods.

\vspace{-2mm}
\subsection
{Sensitivity Analysis}\label{sec:sensitivity}

In Section \ref{sec:beta} we study the sensitivity of the number of allocations to $\beta$,
which is used in conjunction with Min-F1 and Min-F2 allocation methods.
In Section \ref{sec:effect} we study the sensitivity 
of allocation results with respect to the granularity of allocations:
$\rho_{max}$ and $v_{max}$.

\vspace{-3mm}
\subsubsection
{Sensitivity to $\beta$ in Min-F1 and Min-F2}\label{sec:beta}

The parameter $\beta$ emphasizes capacity versus bandwidth utilization 
for Min-F1 and Min-F2 allocation methods.
Higher values of $\beta$ assign more weight to disk capacity utilization,
so that more balanced disk capacity utilizations are achieved.

The following parameters are used in experiments: 
$N=12$ disks, RAID5:RAID1=3:1, 
the fraction of read requests is $r=1$. 
The VA sizes are exponentially distributed 
with $\overline{V}_1 = 256$ MB for RAID1 
and $\overline{V}_5 = 768$ MB for RAID5. 
The access rates for RAID1 are ten times 
the access rates of RAID5: $( \kappa_1= 10 \kappa_5)$. 
$v_{max}$ is set to $ 1/50^{th}$ of the capacity of all disks 
and $\rho_{max}$ is set to $ 1/20 $ of the bandwidth of a single disk. 
We consider bandwidth-bound, balanced, and capacity-bound workloads.

The following conclusions can be drawn from Table \ref{alphaSensF1}.
For balanced and capacity-bound workloads the number of VAs allocated
increases more than 10\% when $\beta$ is increased from zero to one,
but there is little improvement for $\beta > 1$.
The effect on the bandwidth-bound workload is also less significant.
Note that a large value for $\beta$ would result 
in the effect of bandwidth utilizations being ignored.
We have used $\beta = 1.0$ in the experimental results reported in this study.

\begin{table}[t]
\centerline{ \small
\begin{tabular}{|c||c|c|c|c|}
\hline
    $\beta$ & Bandwidth-Bound & Balanced & Capacity-Bound\\
\hline\hline
    $\ge$ 1 & 116/116 & 127/129 & 139/139\\
\hline
    0.5 & 108/116 & 123/129 & 133/139\\
\hline
    0 & 108/108 & 108/108 & 123/123 \\
\hline
\end{tabular}}
\caption{\label{alphaSensF1}\small Sensitivity of $\beta$ in
Min-F1/Min-F2 with RAID5:RAID1=3:1, and $r=1$ in degraded mode.}
\end{table}

\vspace{-3mm}
\subsubsection 
{The Effect of $\rho_{max}$ and $v_{max}$ on Allocations}
\label{sec:effect} 

The parameters $\rho_{max}$ and $v_{max}$ determine the maximum bandwidth 
(or equivalently maximum disk bandwidth utilization) 
and the maximum capacity allocated per VD on a disk drive, respectively.
$\rho_{max}$ is the more critical parameter,
since it restricts the bandwidth utilization of a VA on a disk.
Because VA loads are estimates, restricting the load per VD reduces 
the possibility of a disk becoming overloaded when the VA load 
was underestimated in the allocation phase.
The smaller the values of these two parameters, 
the more VDs are required for the allocation and vice-versa.

The following parameters are used in the experiment:
$N=12$, RAID5:RAID1=1:0, i.e., only RAID5 requests, and $r=1$,
since the load increase in degraded mode is highest for read requests. 
The VA sizes are exponentially distributed 
with $\overline{V}_1 = 256$ MB for RAID1 and $\overline{V}_2 = 768$ MB for RAID5. 
The allocation method Min-F1 is used in this experiment, 
because it is one of the two best methods 
as shown in Section \ref{sec:allocation} with $\beta$ set to one.
We consider the bandwidth-bound, balanced, and capacity-bound workloads.

\vspace{-2mm}
\begin{table}[htp]
\centerline{\small
\begin{tabular}{|c||c|c|c||c|c|c||c|c|c|}
\hline
& \multicolumn{3}{c||}{Bandwidth-bound} & \multicolumn{3}{c||}{Balanced} &
\multicolumn{3}{c|}{Capacity-bound}\\
\cline{1-10} \hline\hline
$\rho_{max}$  & 4\% & 2\% & 1\% & 4\% & 2\% & 1\% & 4\% & 2\% & 1\%  \\
\hline
1/10 & 46   &   46   &  46     & 74   & 93     &  123  & 88   & 104     &  135\\
1/20   & 49   & 49     &  49  & 77   & 96     &  125  & 88   & 104     &  135\\
1/40   & 50   & 50     &  50  & 82   & 99     &  128  & 88   & 104     &  135\\
\hline
\end{tabular}
}
\caption{\label{effects}\small
The effects of $\rho_{max}$ (given in the first column)
and $v_{max}$ (the percentages in the second row) 
on the number of VAs allocations in degraded mode.}
\end{table}

The following conclusions can be drawn from Table \ref{effects}.
$\rho_{max}$ affects the number of VDs for both balanced and bandwidth-bound workloads,
but has no effect on capacity-bound workloads,
since the width is determined by $v_{max}$ rather than $\rho_{max}$.
If we consider $\rho_{max}=1/20$ and a capacity-bound system,
the increase in the number of allocations can be explained 
by the fact that for $v_{max} = 1/25$, $1/50 $, and $1/100$
the mean disk capacity utilization was determined to be
96.5\%, 98.5\%, and 99.2\%, respectively,
while the standard deviation of capacity utilization was quite small in all cases.
The improvement in allocation efficiency is due 
to the reduction in the sizes of allocation requests.
Incidentally, the average RAID5 width was determined to be $4.7$, $5.6$, $7.3$ for 
$\rho_{max} =1/10$,
$\rho_{max} =1/20$,
$\rho_{max} =1/40$, respectively.

\vspace{-2mm}
\subsection
{Clustered RAID5}\label{sec:craid} 

Clustered RAID5 is a method to attain a reduction 
in disk load increase when operating in degraded mode.
It can be used to attain more allocations for 
bandwidth-bound workloads when a sufficient disk capacity is available.  

Given the width of VAs $W \leq N$ and parity group size $G \leq W $, 
we study the effect of the declustering ratio: $\alpha = (G-1)/(W-1)$ 
on the number of allocated VAs with the clustered RAID5 organization.
This experiment could have been carried out with read and write requests, 
since disk loads for RAID5 with and without clustering are given in \cite{Thom05b},
but this discussion is simplified by postulating only read accesses,
since the increase in disk loads in degraded mode is highest in this case 
and can be simply expressed as $1 + \alpha$.
Note that a clustered RAID5 with $G=2$ corresponds to RAID1
and one disk access is required to reconstruct a block on the failed disk.

For a disk drive with capacity $C$ with a workload consisting
of accesses to randomly placed small data blocks,
the {\it Capacity/Bandwidth Ratio (CBR)} is given as: $\gamma_d= C \overline{x}_{SR}$.
where $[\overline{x}_{SR}]^{-1}$ is the maximum disk bandwidth for SR requests. 
{\it Bandwidth/Space Ratio (BSR)}, which is the inverse of CBR, 
has been defined in the context of storage systems for multimedia applications.
The capacity of IBM 18ES disk drives considered in this study is $C=9.17$ GB
and their maximum bandwidth is $1000 / \overline{x}_{SR} \approx 87$ accesses per second, 
so that $\gamma_{d} = 9.17/87 = 0.105$ (GB/second). 
This corresponds to the slope of the diagonal in Figure~\ref{fig:vectors}.
It is desirable for $\gamma_c (\alpha)$ for RAID5 to be close to the disk CBR: 
$\gamma_c (\alpha) \approx \gamma_d$,
such that successive allocation vectors follow the diagonal in Figure~\ref{fig:vectors}.

The RAID5 access rate per GB with a bandwidth-bound workload is $\kappa_5 = 8.5 $ accesses per second.
The capacity bandwidth ratio for a workload 
for clustered RAID5 is given in Table \ref{gamma-1-0}. 
For RAID5 without clustering $\gamma_c(1) < \gamma_d$, 
but $\gamma_c (0.25) \approx \gamma_d$.
\footnote{To illustrate the workings of this table note 
that the bandwidth 15.22 for $\alpha =1$ 
is twice the bandwidth in normal mode, i.e., 7.61.
The load increase for $\alpha = 0.125$ is $7.61 \times 1.125 = 8.56$ GB.
Without clustering the size of the data portion of the VA is  
$ V_D =  0.87 \times 11/12 \approx 0.8 $ GBs.
For $\alpha =0.125$ and $G= 1 + \alpha (W-1) \approx 2.375$.
The size of the clustered VA is then 
$V(0.125) = 0.8 \times (1 + 1/2.375)  \approx 1.14$ GB.} 

\begin{table}[htp]
\centerline{ \small
\begin{tabular}{|c||c|c|c|c|}
\hline
$\alpha$   &   G   &   $Capacity$   &   $Bandwidth$ &   $\gamma_c (\alpha)$ \\
\hline\hline
0.125   &   2   &   1.14    &   8.56    &   0.133    \\
\hline
0.25    &   4   &   1.01    &   9.51    &   0.106    \\
\hline
0.375   &   5   &   0.96    &   10.46    &   0.091    \\
\hline
0.5 &   7   &   0.92    &   11.42    &   0.081    \\
\hline
0.625   &   8   &   0.90    &   12.37    &   0.073    \\
\hline
0.75    &   9   &   0.89    &   13.32   &   0.067    \\
\hline
0.875   &   11  &   0.88    &   14.27   &   0.061    \\
\hline
1   &   12  &   0.87    &   15.22   &   0.057    \\
\hline
\end{tabular}}
\caption{\small \label{gamma-1-0} \small 
Change of capacity/bandwidth ratio ($\gamma_c$) for VAs 
with 8.5 accesses per second per GB versus the declustering ratio 
$\alpha$ in clustered RAID5 with all read requests ($r=1$).}
\end{table}

\vspace{-2mm}
\begin{table}[htp]
\centerline{  \small
\begin{tabular}{|c@{}||c|c|c|}
\hline
$\alpha$    &   $f_r=1.0$ &  $f_r=0.75$ &  $f_r = 0.5$ \\
\hline\hline
0.25 &   108.6 (c)   &   69.4 (b)    &   50.6 (b)\\
\hline
0.5 &   92.0  (b)  & 61.7 (b) & 46.4 (b)\\
\hline
0.75    &   78.8 (b)   & 55.5 (b)    & 42.8 (b)\\
\hline
\end{tabular}}
\caption{\label{alpha-N-degraded} \small 
Number of RAID5 allocations with bandwidth-bound workload 
(8.5 access per second per GB) in degraded mode with $N=12$ disks. 
The (c)/(b) next to the number of allocations 
indicates the capacity/bandwidth reaches the limit first.}
\end{table}


\vspace{-2mm} 
\begin{table}[htp]
\centerline{  \small
\begin{tabular}{|c@{}||c|c|c|}
\hline
$\alpha$    & $f_r=1.0$  &  $f_r=0.75$ & $f_r=0.50$ \\
\hline\hline
0.25 &   94.6 (c)   &   74.8 (b)    &   53.4 (b)\\
\hline
0.5 &   106.3 (c)  & 70.7 (b) & 51.3 (b)\\
\hline
0.75    &   103.9 (b)   & 67.0 (b)    & 49.4 (b)\\
\hline
\end{tabular}}
\caption{\label{alpha-W-degraded} \small 
Number of RAID5 allocations with bandwidth-bound workload with 
$\kappa_5 = 8.5$ accesses per sec/GB in degraded mode with $N=12$ disks
The (c)/(b) next to the number of allocations means 
the capacity/bandwidth reaches the limit first.}
\end{table}


\begin{table}[htp]
\centerline{ \small
\begin{tabular}{|c||c|c|c|c|c|c|}
\hline
&   \multicolumn{2}{c|}{RAID5(width=$W$)}& \multicolumn{3}{c|}{Clustering+$W$}\\
\cline{2-6}
                 & Rel \# & Avg Width &  Rel \# & Avg Width & Avg $\gamma$ \\
\hline\hline
    $f_r= 1.0$  & 1.45 & 5.8 & 1.57 & 3.9 & 0.106\\
\hline
    $f_r= 0.75$  & 1.30 & 5.7 & 1.49 & 2.03 & 0.086\\
\hline
    $f_r =0.50$  & 1.22 & 5.6 &  1.34 & 2.8 & 0.06\\
\hline
\end{tabular}}
\renewcommand{\baselinestretch}{1}\small
\caption{\small \label{results-1-0}\small
Comparison of the relative number of allocations
with and without clustering in degraded mode with respect to setting $W=N$.}
\end{table}

We make allocation experiments with $\alpha=0.25, 0.5,  0.75$.
The following conclusions can be drawn from Tables~\ref{alpha-N-degraded} and 
\ref{alpha-W-degraded}, 

\begin{enumerate}

\vspace{-2mm}
\item
If the disk bandwidth is the bottleneck resource 
then more allocations can be made with smaller values of $\alpha$,
since the bandwidth requirement per VA is lower in degraded mode.

\vspace{-2mm}
\item
As $\alpha$ decreases the system reverts from bandwidth to capacity bound.
This is because as $G= \alpha (N-1) +1$ gets smaller, the disk capacity overhead $1/G$ gets larger.
The number of allocations increases, until the disk capacity becomes the bottleneck resource.

\vspace{-2mm}
\item
If disk capacity is the bottleneck resource
then more allocations can be made with larger values of $\alpha$,
because the capacity overhead per VA is lowest for $G=N$.

\end{enumerate}

Finally, we use the general configuration for all experiments
and a bandwidth-bound workload with the Min-F1 method and $\beta = 1$.
For each allocated VA $\alpha$ is set such that VAs CBR $\gamma_c$ is close to $\gamma_d$.
Once the value of $\alpha$ is determined, the new width will be calculated accordingly.
It can be concluded from Table~\ref{results-1-0} that 
clustered RAID5 arrays increase the number of VA 
allocations significantly when all requests are reads ($r=1$),
but clustering has less effect for higher fractions of write requests.

\section{CONCLUSIONS}\label{sec:conclusions} 

We have described HDA, argued about its advantages,
and used a synthetic workload to experimentally compare several single-pass data allocation methods for HDA.
The allocation efficiency is determined primarily by the number of allocated VAs,
disk bandwidth and capacity utilizations, and by the balancedness of allocations.

While only RAID1 and RAID5 are considered in the experimental study, 
we have provided the methodology for calculating the load for other RAID levels.
These analytic results can be used in further studies of HDA.

The two allocation methods: Min-F1 and Min-F2, 
which take into account both disk access bandwidth and capacity, 
outperform methods such as Worst-Fit and Best-Fit, 
which only take into account disk bandwidth.
Methods in the latter category outperform methods such as Round-Robin and Random,
which do not take into account either disk bandwidth or capacity.

Experimental results that the better data allocation methods result in disks 
with fully utilized bandwidth, capacity, or both bandwidth and capacity
for bandwidth-bound, capacity-bound, and balanced workloads, respectively.
More complex data allocation methods would be required
to deal with larger allocation requests with respect to disk bandwidth and capacity.

We have investigated the sensitivity of allocations to $\beta$, 
emphasizing disk space utilization, and $\rho_{max}$ and $v_{max}$,
which limit the bandwidth and capacity utilization per VD. 
The sensitivity to $\beta$ will be less important if the increase in
disk capacities exceeds dataset sizes. 
Experimental results show that a smaller $\rho_{max}$ 
results in an increased number of VA allocations for bandwidth-bound VAs,
which is consistent with the fact that smaller allocations yield a higher bin-packing efficiency.

Clustered RAID5 is a means of decreasing 
the increased load on surviving disks when disk failures occur.
An experimental approach was used to quantify the increase in number of allocations
and the optimal value for the declustering ratio ($\alpha$).



Intermixed bandwidth- and capacity-bound requests 
can benefit from an adaptive allocation method, 
which monitors the end point of allocations over time 
with respect to the disk vector in Figure~\ref{fig:vectors}
and applies declustering to bandwidth-bound 
and data compression to capacity-bound allocation requests.

Methods to reallocate VAs whose load was incorrectly estimated 
or varied significantly over time are required for system maintenance.
The disk cooling algorithm proposed in \cite{ScWZ98},
which balances disk loads by reallocating datasets according to their ``heat'', i.e., access rate,
is relevant to this discussion.
Similar capabilities have been developed for storage migration for VMware's virtual machines (vMotion).

Overload control, especially in degraded mode,
can be applied to throttle access to VAs for less critical applications
to ensure satisfactory performance for more critical ones. 
There is the more general issue of attaining {\it Quality of Service (QoS)} 
for different applications sharing disk space \cite{Bor+97}. 

An alternate formulation as a bin-packing problem is to start with a fixed number of VAs to be allocated,
with the goal to minimize the number of utilized HDAs.
Processing allocations in decreasing order of their load will yield improved results. 

Rebuild processing is a systematic reconstruction of a failed disk, 
which is accomplished by reading the contents of surviving disks to reconstruct missing data.
Rebuild processing can be carried out using:
(1) a spare disk, in the case of HDA an additional VD need be allocated;
(2) spare areas as in distributed sparing \cite{ThMe97},
but this approach does not make sense in the context of HDA,
since it would require larger VA sizes;
(3) overwriting check strips in the case of restriping,
so that a RAID5 would be converted to a RAID0 and a RAID6 to RAID5 \cite{ThTH12};
(4) parity sparing by combining same-sized RAID5 arrays \cite{ThBl09}.
Methods (2) and (3) have the advantage of balancing disk loads
and not requiring extra space allocation 

Simple instances of HDA performance analysis are given in Section~\ref{sec:justification}.
Performance evaluation of HDA via analysis,
simulation, or measurements and benchmarking of a prototype remain an area of further investigation.
Benchmarks developed by the {\it Transaction Processing Council (TPC)}
\footnote{\url{http://www.tpc.org}.} 
can be used to determine the viability of HDA by developing a credible prototype.
Analytic methods for performance evaluation methods for RAID5 \cite{ThFH07}
and RAID1 \cite{ThXu08} can be extended to HDA operating in normal, degraded, and rebuild modes.


{\large \bf Abbreviations:} 
{\bf BM:} Basic Mirroring.                 
{\bf DAC:} Disk Array Controller.          
%
%
{\bf HDD:} Hard Disk Drive.                
%
%
%
{\bf $k$DFT:} $k$ Disk Failure Tolerant.     
{\bf LSA:} Log-Structured Array.             
{\bf LSE:} Latent Sector Error.              
%
%
%
{\bf NVS:} Non-Volatile Storage.
%
%
%
{\bf RMW:} Read-Modify-Write.                 
{\bf SR/SW:} Single Read/Write.               
%
%
%
{\bf VA/VD:}  Virtual Array/Disk.                     
%
%
%
{\it XOR:} eXclusive OR.

\clearpage

\clearpage
\section*{Appendix: Related Work}\label{sec:related}

``Selection of RAID levels and stripe characteristics 
based on application characteristics was and is a black art'' 
\footnote{\url{http://www.openmpe.com/cslproceed/HPW98CD/3000/3354/3354.htm}.}
HP's AutoRAID is then a solution to this problem \cite{WGSS96}. 
It is a hierarchical disk array with two levels: 
RAID1 at the higher level and RAID5/{\it Log-Structured Array (LSA)}, at the lower level.
The commonly adopted inclusion method for multilevel CPU caches is not followed \cite{HePa07}, 
i.e., RAID1 data is not replicated by RAID5.
AutoRAID may be initially filled with data formatted as RAID1, 
but as disk space is exhausted, mirrored data is demoted to RAID5/LSA format,
which obviates the small write penalty by writing data in full stripes, 
so that check strips are computed on the fly \cite{Che+94,ThBl09}. 
Excessive disk activity resulting in thrashing is potentially a problem 
when the size of the active working set exceeds the space available for RAID1. 

A succession of tools to create self-configuring and self-managing storage systems at HP: 
{\it Forum} \cite{Bor+97} and {\it Minerva} \cite{Alv+01}
culminated in {\it Disk Array Designer (DAD)} \cite{And+05}.
Minerva is a suite of tools for automated design of large storage systems.
Given the descriptions of the workload and the capabilities of storage devices,
the output is an {\it assignment}.
It is assumed that disk arrays provide {\it Logical UNits (LUNs)}, which may be RAID5 or RAID1 arrays.
The workload description is in the form of {\it stores} (chunks of data) and {\it streams}, 
which are accesses to the stores.
Given the RAID level Minerva carries out the following steps:
(i) array allocation, (ii) array configuration, and (iii) store assignment. 
DAD uses best-fit bin-packing with randomization and backtracking.
It can be utilized to improve the current storage configuration 
as part of an automated storage management system.
The data allocation experiments in \cite{And+05}
are based on workload characterization in several studies at HP.

Oracle's {\it Stripe and Mirror Everywhere (SAME)} paradigm 
is obviously not the best storage scheme in all cases \cite{And+05}.
RAID level selection: RAID1/0 versus RAID5 is addressed in \cite{And+02}.
{\it Tagging-based} and {\it solver-based} approaches are considered.
Tagging may be {\it rule-based}, e.g., rules of thumb, 
or {\it model based}, which selects the RAID level minimizing the I/O rate.
The solver-based approach has two variants:
{\it partially adaptive} and {\it fully adaptive}.
The former does not allow RAID levels to be reassigned and the latter does.

An analytic approach to select RAID levels based on access characteristics 
of allocation requests is proposed in \cite{ThXu11}.
This study postulates a RAID5 that allows in-place updating of small blocks
and full stripe updating of large blocks as in the case of RAID5/LSA.
The RAID level yielding the lower load: RAID1 or RAID5, is selected.
The current paper does not rely on this work 
and assumes that the RAID level for VAs is specified a priori.

RAID1 is the appropriate RAID level for small write requests,
since it incurs less overhead than RAID$(4+k), k \geq 1$.
RAID1/0 is a generalization of {\it Basic Mirroring (BM)} with multiple pairs of mirrored disks.
When  a disk fails, the read load of its pair is doubled, but this is not so for the data layouts 
where data of each disk is replicated on more than one disk \cite{ThXu08}.
Space requirement preclude the discussion of such RAID1 organizations.

RAID5 is more efficient than RAID1 in dealing with large datasets.
In addition to savings in disk space,
large files can be written more efficiently as {\it full stripe writes}.
This allows the few check strips to be calculated on-the-fly,
while RAID1 would require writing the dataset twice.
The strips of a RAID5 file can be read in parallel speeding up the access.
In the case of EMC's Centera file system for archival storage
small files are stored as RAID1 and large files as RAID5. 
\footnote{\url{http://www.emc.com/products/family/emc-centera-family.htm}.} 

Heterogeneity at both the RAID and disk level was investigated in \cite{ThHa05}.
RAID1 and RAID5 ``containers'' with prespecified widths are preallocated when space in either category is exhausted.
VDs of VAs tagged as RAID1 or RAID5 with prespecified loads 
are allocated in the containers taking into account disk utilizations, as well as container capacity.
 


REO (RAID Engine and Optimizer) works for any XOR-based erasure code
and any combination of sector or disk failures \cite{KeHH07}.
The emphasis of this study is on firmware reuse and reducing complexity.
Benchmarking results with various workloads indicate
that REO may attain modest improvements over existing RAID implementations.

The {\it Redundant Array of Independent Filesystems (RAIF)} 
provides RAID support with different levels at the per file level. 
There is heavy reliance on stackable file systems,
which provide flexibility in composition and development.

The Panasas PFS (parallel file system) stores striped files 
as RAID1 or RAID5, depending on their size \cite{Wel+08}.
The four PFS advantages:
(i) Scalable computing of parities for the files by clients.
(ii) Capability for end-to-end data integrity checking.
(iii) Parallel rebuild capability for files.
(iv) Limiting unrecoverable faults to individual files. 
The initial data placement is uniform random.
{\it Passive balancing} is used to place new data into empty nodes,
while {\it active balancing} moves existing objects 
from one node to another to eliminate hotspots.





{\bf Alexander Thomasian} has a PhD in Computer Science from UCLA.
He spent half of his career in academia:
Case Western Reserve Univ. (CWRU),
Univ. Southern Calif. (USC), 
and New Jersey Institute of Technology (NJIT);
and the other half at IBM's T. J. Watson Center and Almaden Research Center (ARC).
He continued research in storage systems started at ARC at NJIT with funding from NSF,
Hitachi Global Storage Technologies (HGST), and AT\&T Research.
He spent a year at Shenzhen Institutes of Advanced Technology 
as an Outstanding Visiting Scientist to the Chinese Academy of Sciences
and taught at the American University in Armenia in Yerevan on a Fulbright Scholarship.
He has authored key papers on analysis and synthesis of concurrency control methods.
He has utilized a combination of clustering and SVD to high dimensional datasets
and evaluated nearest neighbor queries on indices on such data.
He was an area of editor of IEEE Trans. Parallel and Distributed Systems
and on the program committees of numerous conferences.
He has authored over 120 journal and conference papers, a dozen book chapters, and four patents.
 
{\bf Jun Xu} holds a BS degree in economics from 
Shanghai Int'l Business and Economics Univ., 
an MBA from Univ. Baltimore in Maryland, 
and a PhD in Computer Science from NJIT. 
He is a Senior Developer at Six Financial Information in New York City

\end{document}